\documentclass{nature}

\usepackage{cite}
\usepackage{graphicx}
\usepackage{paralist}
\usepackage[usenames,dvipsnames]{color}
\usepackage[T1]{fontenc}
\usepackage{times}
\usepackage{inconsolata}
\usepackage{framed}
\usepackage[utf8]{inputenc}

\newcounter{infobox}[section]
\usepackage{mdframed}

\newenvironment{infobox}[1]
  {\refstepcounter{infobox} \ \\ ***** START BOX {\theinfobox}:\\ \ \\{#1}}
  {\ \\***** END BOX {\theinfobox}}

\usepackage{cite}
\makeatletter\def\@cite#1#2{\nolinebreak$^{\mbox{\scriptsize\,#1\if@tempswa,#2\fi}}$}\makeatother

\usepackage{enumerate}

\definecolor{darkblue}{cmyk}{1,1,0,.2}
\usepackage[colorlinks=true,citecolor=darkblue,linkcolor=darkblue,urlcolor=darkblue]{hyperref}

\newcommand{\Timing}[1]{\textsl{\textcolor{WildStrawberry}{\emph{\ TIMING}:#1}}}

\newcommand{\edgeRroman}[2][]{\vspace{3.5ex}\noindent\refstepcounter{step}\textbf{{\thestep}. #1}{#2}\\\nopagebreak\@afterindentfalse}

\newcounter{step}
\makeatletter
\newcommand{\Step}[2][]{\vspace{3.5ex}\noindent\refstepcounter{step}\textbf{{\thestep}. #1}{#2}\\\nopagebreak\@afterindentfalse}
\makeatother

\newcommand{\fileformat}[1]{\MakeUppercase{\textsf{#1}}}
\newcommand{\tool}[1]{\texttt{#1}}

\newcommand{\Rfunction}[1]{{\color{Cerulean} \texttt{#1}}}
\newcommand{\Rclass}[1]{{\color{Cerulean} \texttt{#1}}}
\newcommand{\Robject}[1]{\texttt{#1}}

\title{Count-based differential expression analysis of RNA sequencing data using \tool{R} and \tool{Bioconductor}}

\author{Simon Anders$^1$, Davis J. McCarthy$^{2,3}$, Yunshen Chen$^{4,5}$, Michal Okoniewski$^6$, Gordon K.\ Smyth$^{4,7}$, Wolfgang Huber$^{1,*}$ \& Mark D. Robinson$^{8,9,*}$}

\usepackage{Sweave}
\begin{document}


\maketitle

\begin{affiliations}
 \item Genome Biology Unit, European Molecular Biology Laboratory, Mayerhofstrasse 1, 69117 Heidelberg, Germany
 \item Department of Statistics, University of Oxford, 1 South Parks Road, Oxford, OX1 3TG, United Kingdom
 \item Wellcome Trust Centre for Human Genetics, University of Oxford, Roosevelt Drive, Oxford, OX3 7BN, United Kingdom
 \item Bioinformatics Division, Walter and Eliza Hall Institute, 1G Royal Parade, Parkville, Victoria 3052, Australia
 \item Department of Medical Biology, University of Melbourne, Victoria 3010, Australia
 \item Functional Genomics Center UNI ETH Zurich, Winterthurerstrasse 190, CH-8057, Switzerland
 \item Department of Mathematics and Statistics, University of Melbourne, Victoria 3010, Australia
 \item Institute of Molecular Life Sciences, University of Zurich, Winterthurerstrasse 190 CH-8057 Zurich, Switzerland
 \item SIB Swiss Institute of Bioinformatics, University of Zurich, Zurich, Switzerland
\end{affiliations}


$^{*}$ Correspondence and requests for materials should be addressed to M.D.R. or W.H.  (email: mark.robinson@imls.uzh.ch, whuber@embl.de) \ \\

\begin{abstract}
  RNA sequencing (RNA-seq) has been rapidly adopted for the profiling of transcriptomes in many areas of biology,
  including studies into gene regulation, development and disease.
  Of particular interest is the discovery of differentially expressed genes across different conditions (e.\,g., tissues, perturbations), 
  while optionally adjusting for other systematic factors that affect the data collection process. 
  There are a number of subtle yet critical aspects of these analyses, such as read counting,
  appropriate treatment  of biological variability, quality control checks and
  appropriate setup of statistical modeling.  Several variations have been
  presented in the literature, and there is a need for guidance on current
  best practices. This protocol presents a ``state-of-the-art''
  computational and statistical RNA-seq differential expression analysis workflow
  largely based on the free open-source \tool{R} language and \tool{Bioconductor} software and in particular, 
  two widely-used tools \tool{DESeq} and \tool{edgeR}.
  Hands-on time for typical small experiments (e.\,g., 4-10 samples) can be <1 hour, with computation
  time <1 day using a standard desktop PC.
\end{abstract}

\section*{INTRODUCTION}

\noindent \textbf{Applications of the protocol.}  The RNA sequencing (RNA-seq) platform\cite{Mortazavi2008,Wang2009} addresses a multitude of applications, including relative expression analyses, alternative splicing, discovery of novel transcripts and isoforms, RNA editing, allele-specific expression and the exploration of non-model organism transcriptomes.



Typically, tens of millions of sequences (``reads") are generated, and these, across several samples, form the starting point of this protocol.  An initial and fundamental analysis goal is to identify genes that change in abundance between conditions.  In the simplest case, the aim is to compare expression levels between two conditions, e.\,g., stimulated versus unstimulated or wild-type versus mutant.  More complicated experimental designs can include additional experimental factors, potentially with multiple levels (e.\,g., multiple mutants, doses of a drug or time points) or may need to account for additional covariates (e.\,g.\, experimental batch or sex) or the pairing of samples (e.\,g., paired tumour and normal tissues from individuals).  A critical component of such an analysis is the statistical procedure used to call differentially expressed genes. This protocol covers two widely-used tools for this task: \tool{DESeq}\cite{Anders2010} and \tool{edgeR}\cite{Robinson2007,Robinson2008,Robinson2010a,McCarthy2012}, both available as packages of the Bioconductor software development project\cite{Gentleman2004}.

\noindent Applications of these methods to biology and biomedicine are many-fold.  The methods described here are general and can be applied to situations where observations are counts (typically, hundreds to tens of thousands of features of interest) and the goal is to discover of changes in abundance.  RNA-seq data is the typical use-case (e.g., \cite{Zemach2013,Lam2013}), but many other differential analyses of counts are supported\cite{Ross-Innes2012,Robinson2012}.  For RNA-seq data, the strategy taken is to count the number of reads that fall into annotated genes and perform the statistical analysis on the table of counts to discover quantitative changes of expression levels between experimental groups.  This counting approach is direct, flexible and can be used for many types of count data beyond RNA-seq, such as comparative analysis of immunoprecipitated DNA\cite{Ross-Innes2012,Robinson2012,Vanharanta2013,Samstein2012} (e.\,g.\ ChIP-seq, MBD-seq;\cite{Ross-Innes2012,Robinson2012}), proteomic spectral counts\cite{Johnson2012} and metagenomics data.

\begin{figure}
\centering
\includegraphics[width=.8\textwidth]{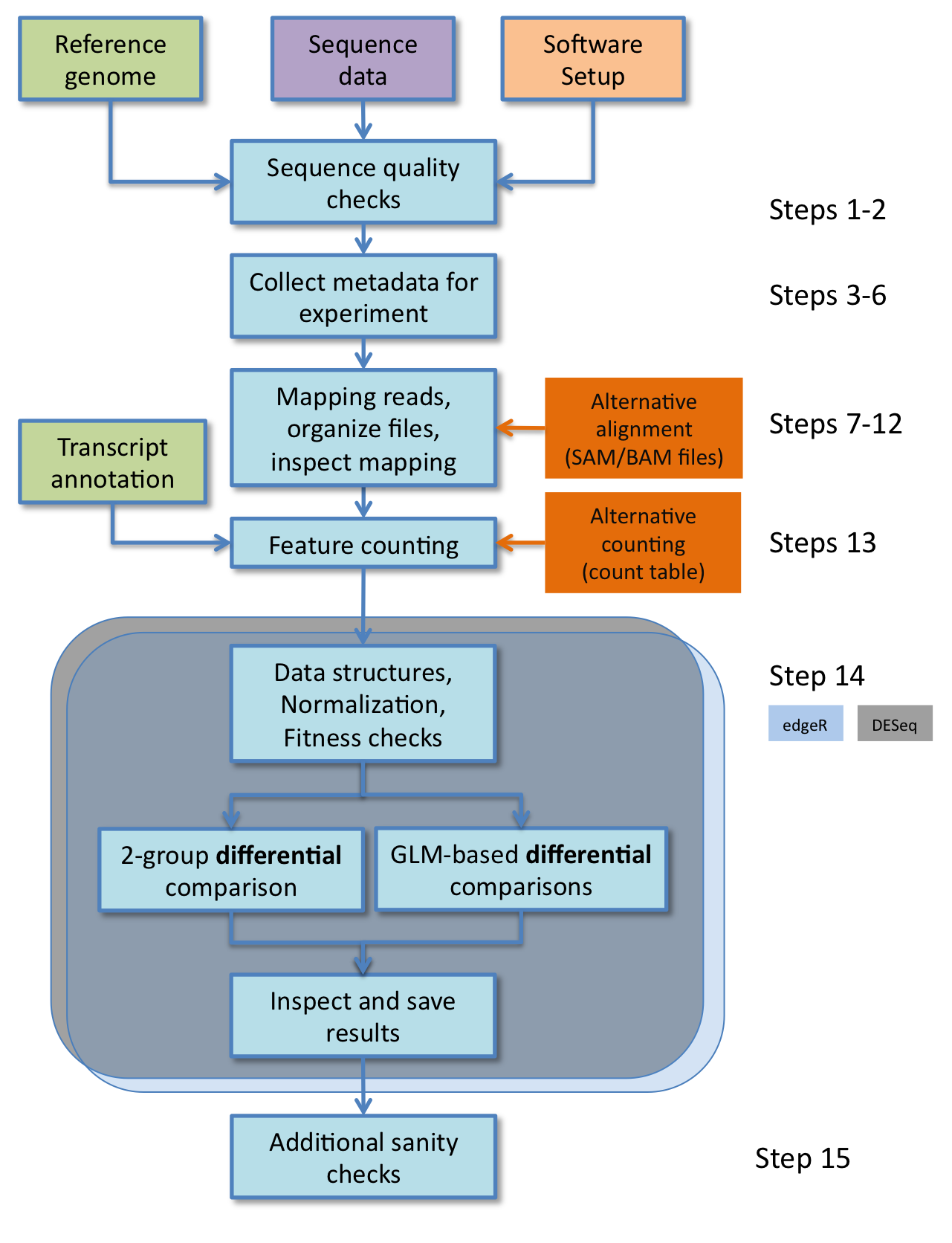}
\caption{Count-based differential expression pipeline for RNA-seq data using \tool{edgeR} and/or \tool{DESeq}.  Many steps are common to both tools, while the specific commands are different (Step \ref{DEanalysis}).  Steps within the \tool{edgeR} or \tool{DESeq} differential analysis can follow two paths, depending on whether the experimental design is \emph{simple} or \emph{complex}.  Alternative entry points to the protocol are shown in orange boxes. }
\label{pipeline-fig}
\end{figure}

\noindent \textbf{Development of the protocol.}  Figure~\ref{pipeline-fig} gives the overall sequence of steps, from read sequences to feature counting to the discovery of differentially expressed genes, with a concerted emphasis on quality 
checks throughout. After initial checks on sequence quality, reads are mapped to a reference genome with a splice-aware aligner\cite{Fonseca2012}; up to this point, this Protocol\cite{Robinson2010a,Anders2010} is identical to many other pipelines (e.\,g., TopHat and Cufflinks\cite{Trapnell2012}).  From the set of mapped reads and either an annotation catalog or an assembled transcriptome, features, typically genes or transcripts, are counted and assembled into a table (rows for features and columns for samples).  The statistical methods, which are integral to the differential expression discovery task, operate on a feature count table. Before the statistical modeling, further quality checks are encouraged to ensure that the biological question can be addressed.  For example, a plot of sample relations can reveal possible batch effects and can be used to understand the similarity of replicates and overall relationships between samples.  After the statistical analysis of differential expression, a set of genes deemed to be differentially expressed or the corresponding statistics can be used in downstream interpretive analyses in order to confirm or generate further hypotheses.

Replication levels in designed experiments tend to be modest, often not much more than two or three. As a result, there is a need for statistical methods that perform well in small-sample situations.  The low levels of replication rule out, for all practical purposes, distribution-free rank- or permutation-based methods.  Thus, for small to moderate sample sizes, the strategy employed is to make formal distributional assumptions about the data observed.  The advantage of parametric assumptions is the ability, through the wealth of existing statistical methodology, to make inferences about parameters of interest (i.\,e., changes in expression).  For genome-scale count data including RNA-seq, a convenient and now well-established approximation is the negative binomial (NB) model (see Box \ref{NBbox} for further details), which represents a natural extension of the Poisson model (i.\,e., mixture of Gamma-distributed rates) that was used in early studies\cite{Bullard2010}; importantly, Poisson variation can only describe technical (i.\,e., sampling) variation. 

For the analysis of differential expression, this protocol focuses on \tool{DESeq} and \tool{edgeR}, which implement general differential analyses based on the NB model.  These tools differ in their ``look-and-feel'' and estimate the dispersions somewhat differently but offer overlapping functionality (See Box~\ref{edgeRdeseqdiffbox}).

\noindent \textbf{Variations and extensions of the protocol.} This protocol presents a workflow built from a particular set of tools, but it is modular and extensible, so alternatives that offer special features (e.\,g., counting by allele) or additional flexibility (e.\,g., specialized mapping strategy), can be inserted as necessary.  Figure~\ref{pipeline-fig} highlights straightforward alternative entry points to the protocol (orange boxes).  The count-based pipeline discussed here can be used in concert with other tools. For example, for species without an available well-annotated genome reference, Trinity\cite{Grabherr2011} or other assembly tools can be used to build a reference transcriptome;  reads can then be aligned and counted, followed by the standard pipeline for differential analysis\cite{Siebert2011}.  Similarly, to perform differential analysis on novel genes in otherwise annotated genomes, the protocol could be expanded to include merged per-sample assemblies (e.\,g.\ \tool{cuffmerge} within the \tool{cufflinks} package\cite{Trapnell2009a,Trapnell2010,Trapnell2012}) and used as input to counting tools.

The focus of this protocol is gene-level differential expression analysis.  However, biologists are often interested in analyses beyond that scope, and many possibilities now exist, in several cases as extensions of the count-based framework discussed here.  Here, the full details of such analyses are not covered, and only a sketch of some promising approaches is made.  First, an obvious extension to gene-level counting is exon-level counting, given a catalog of transcripts.  Reads can be assigned to the exons that they aligned to, and these assignments be counted.  Reads spanning exon-exon junctions can be counted at the junction level.  The \tool{DEXSeq} package uses a GLM that tests whether particular exons in a gene are preferentially used in a condition, over and above changes in gene-level expression. In \tool{edgeR}, a similar strategy is taken, except that testing is done at the gene-level, effectively asking whether the exons are used proportionally across experiment conditions, in the context of biological variation.

\noindent \textbf{Comparison to other methods.}  Many tools exist for differential expression of counts, with slight variations of the method demonstrated in this protocol; these include, among others, \tool{baySeq}\cite{Hardcastle2010}, \tool{BBSeq}\cite{Zhou2011}, \tool{NOISeq}\cite{Tarazona2011} and \tool{QuasiSeq}\cite{Lund2012}.  The advantages and disadvantages of each tool are difficult to elicit for a given dataset, but simulation studies show that \tool{edgeR} and \tool{DESeq}, despite the influx of many new tools, remain among the top performers\cite{Soneson2013}.

The count-based RNA-seq analyses presented here consider the \emph{total} output of a locus, without regard to the isoform diversity that may be present. This is of course a simplification.  In certain situations, gene-level count-based methods may not recover true differential expression when some isoforms of a gene are up-regulated and others are down-regulated\cite{Lareau2007,Trapnell2012}. Extensions of the gene-level count-based framework to differential exon usage are now available (e.\,g., DEXSeq\cite{Anders2012}; discussed below).  Recently, approaches have been proposed to estimate transcript-level expression and build the uncertainty of these estimates into a differential analysis at the transcript-level (e.\,g., \tool{BitSeq}\cite{Glaus2012}). Isoform deconvolution coupled with differential expression (e.\,g., \tool{cuffdiff}\cite{Trapnell2009a,Trapnell2010,Trapnell2012}) is a plausible and popular alternative, but in general, isoform-specific expression estimation remains a difficult problem, especially if sequence reads are short, if genes whose isoforms overlap substantially should be analysed, or unless very deeply sequenced data is available. At present, isoform deconvolution methods and transcript-level differential expression methods only support two-group comparisons.  In contrast, counting is straightforward, regardless of the configuration and depth of data and arbitrarily complex experiments are naturally supported through GLMs (see Box~\ref{counting-box} for further details on feature counting).  Recently, a flexible Bayesian framework for the analysis of ``random'' effects in the context of GLM models and RNA-seq count data was made available in the \tool{ShrinkSeq} package\cite{VanDeWiel2013}.  As well, count-based methods that operate at the \emph{exon} level, which share the same statistical framework, as well as flexible coverage-based methods have become available to address the limitations of gene-level analyses\cite{Blekhman2010,Anders2012,Okoniewski2011}. These methods give a direct readout of differential exons, genes whose exons are used unequally, or non-parallel coverage profiles, all of which reflect a change in isoform usage.

\noindent \textbf{Scope of this protocol.}
The aim of this Protocol is to provide a concise workflow for a standard analysis, in a
complete and easily accessible format, for new users to the field or to \tool{R}. 
We describe a specific, but very common analysis task, namely the analyis of 
an RNA-Seq experiment comparing two groups of samples that differ in their experimental treatment, and also cover one common complication, namely the need to account for
a \emph{blocking} factor. 

In practice, users will need to adapt this pipeline to account for the circumstances of their experiment. Especially, more complicated experimental designs will require further
considerations not covered here. Therefore, we emphasize that this Protocol is not meant to replace the existing user guides, vignettes and online documentation for the packages and functions described. These provide a large body of information that is helpful to tackle tasks that go beyond the single standard workflow presented here. 

In particular, \tool{edgeR} and \tool{DESeq} have extensive users guides, downloadable from \url{http://www.bioconductor.org}, that cover a wide range of relevant topics.  Please consult these comprehensive resources for further details.  Another  rich resource for answers to commonly asked questions is the \tool{Bioconductor} mailing list (\url{http://bioconductor.org/help/mailing-list/}) as well as online resources such as \url{seqanswers.com}, \url{stackoverflow.com} and \url{biostars.org}.


\noindent \textbf{Multiple entry points to the protocol.}  As mentioned, this protocol is modular, in that users can use an alternative aligner, or a different strategy (or software package) to count features.  Two notable entry points (See orange boxes in Figure~\ref{pipeline-fig}) for the protocol include starting with either: i) a set of \fileformat{SAM}/\fileformat{BAM} files from an alternative alignment algorithm; ii) a table of counts.  With \fileformat{SAM}/\fileformat{BAM} files in hand, users can start at Step~\ref{counting}, although it is often invaluable to carry along metadata information (Steps~\ref{meta-start}-\ref{meta-end}), post-processing the alignment files may still be necessary (Step~\ref{faffingbams}) and spot checks on the mapping are often useful (Step~\ref{igv-start}-\ref{igv-end}).  With a count table in hand, users can start at Step~\ref{DEanalysis}, where again the metadata information (Steps~\ref{meta-start}-\ref{meta-end}) will be needed for the statistical analysis.  For users that wish to learn the protocol using the data analyzed here, Supplementary File 1 gives an archive containing: the intermediate \fileformat{COUNT} files used, a collated count table (\Robject{counts}) in \fileformat{CSV} format, the metadata table (\Robject{samples}) in \fileformat{CSV} format and the \fileformat{CSV} file that was downloaded from the NCBI's Short Read Archive.

\noindent \textbf{Experimental design considerations} \\
\textbf{Replication.}  Some of the early RNA-seq studies were performed
without biological replication. If the purpose of the experiment is to make a general
statement about a biological condition of interest (in statistical parlance, a
population), for example, the effect of treating a certain cell line with a particular
drug, then an experiment without replication is insufficient. Rapid developments in
sequencing reduce technical variation but cannot possibly eliminate biological
variability\cite{Hansen2011}.  Technical replicates are suited to studying properties of
the RNA-seq platform\cite{Fonseca2012}, but they do not inform about the inherent
biological variability in the system or the reproducibility of the biological result, for
instance, its robustness to slight variations in cell density, passage number, drug
concentration or media composition. In other words, experiments without biological
replication are suited to make a statement regarding one particular sample that existed on
one particular day in one particular laboratory, but not whether anybody could reproduce
this result. When no replicates are
available, experienced analysts may still proceed, using one of the following options: i) a descriptive
analysis with no formal hypothesis testing; ii) selecting a dispersion value based on past
experience; iii) using housekeeping genes to estimate variability over all samples in the
experiment.

In this context, it is helpful to remember the distinction between \emph{designed experiments} in which a well-characterized system (e.\,g., a cell line or a laboratory mouse strain) undergoes a fully controlled experimental procedure with minimal unintended variation; and \emph{observational studies}, in which samples are often those of convenience (e.\,g., patients arriving at a clinic) and have been subject to many uncontrolled environmental and genetic factors.  Replication levels of two or three are often a practicable compromise between cost and benefit for designed experiments, whereas for observational studies typically much larger group sizes (dozens or hundreds) are needed to reliably detect biologically meaningful results.

\noindent \textbf{Confounding factors.}  In many cases, data are collected over time.  In this situation, researchers should be mindful of factors that may unintentionally confound their result (e.\,g., batch effects), such as changes in reagent chemistry or software versions used to process their data\cite{Leek2010}. Users should make a concerted effort to: i) reduce confounding effects through experimental design (e.\,g., randomization, blocking\cite{Auer2010}); ii) keep track of versions, conditions (e.\,g., operators) of every sample, in the hope that these factors (or, surrogates of them) can be differentiated from biological factor(s) of interest in the downstream statistical modeling.  In addition, there are emerging tools available that can discover and help eliminate unwanted variation in larger datasets\cite{Gagnon-Bartsch2011,Leek2007}, although these are relatively untested for RNA-seq data at present.



\noindent \textbf{Software implementation.} There are advantages to using a small number of software platforms for such a workflow, and these include simplified maintenance, training and portability. In principle, it is possible to do all computational steps in \tool{R} and \tool{Bioconductor}; however, for a few of the steps, the most mature and widely-used tools are outside \tool{Bioconductor}. Here, \tool{R} and \tool{Bioconductor} are adopted to tie together the workflow and provide data structures, and their unique strengths in workflow components are leveraged, including statistical algorithms, visualization and computation with annotation databases. Another major advantage of an \tool{R}-based system, in terms of achieving best practices in genomic data analysis, is the opportunity for an interactive analysis whereby spot checks are made throughout the pipeline to guide the analyst.  In addition, a wealth of tools is available for exploring, visualizing and cross-referencing genomic data.  Although not used here directly, additional features of \tool{Bioconductor} are readily available that will often be important for scientific projects that involve an RNA-seq analysis, including access to many different file formats, range-based computations, annotation resources, manipulation of sequence data and visualization.

\noindent In what follows, all Unix commands run at the command line appear as:
\begin{verbatim}
my_unix_command
\end{verbatim}

\noindent whereas \tool{R} functions in the text appear as \Rfunction{myFunction}, and
(typed) \tool{R} input commands and output appear as blue and orange, respectively: 

\begin{Schunk}
\begin{Sinput}
> x = 1:10
> median(x)
\end{Sinput}
\begin{Soutput}
[1] 5.5
\end{Soutput}
\end{Schunk}

\noindent Note that in \tool{R}, the operators \verb|=| and \verb|<-| can both be used
for variable assignment (i.\,e., \verb|z = 5| and \verb|z <- 5| produce the same result, a
new variable \Robject{z} with a numeric value). In this
Protocol, we use the \verb|=| notation; in other places, users may also see the \verb|<-|
notation.

\noindent File formats are denoted as \fileformat{PDF} (i.\,e., for Portable Document Format).

\noindent \textbf{Constructing metadata table (Steps~\ref{meta-start}-\ref{meta-end}).} In general, it is recommended to start
from a sample metadata table that contains sample identifiers,
experimental conditions, blocking factors and file names.  In our
example, we construct this table from a file downloaded from the Short Read Archive (SRA; See Download the example data).  
Users will often obtain a similar table from a local laboratory information
management system (LIMS) or sequencing facility and can adapt this strategy to their own
data sets.

\noindent \textbf{Mapping reads to reference genome (Steps~\ref{mapping-start}-\ref{mapping-end}).}  In the protocol, \tool{R} is used to tie the pipeline together (i.\,e., loop through the set of samples and construct the full \tool{tophat2} command), with the hope of reducing typing and copy-and-paste errors.  Many alternatives and variations are possible: users can use \tool{R} to create and call the \tool{tophat2} commands, or just to create the commands (and call \tool{tophat2} independently from a Unix shell), or assemble the commands manually independent of \tool{R}.  \tool{tophat2} creates a directory for each sample with the mapped reads in a \fileformat{BAM} file, called \emph{accepted\_hits.bam}.  Note that \fileformat{BAM} (Binary Alignment Map) files, and equivalently \fileformat{SAM} (Sequence Alignment/Map; an uncompressed text version of \fileformat{BAM}) are the {\em de facto} standard file for alignments.  Therefore, alternative mapping tools that produce \fileformat{BAM}/\fileformat{SAM} files could be inserted into the protocol at this Step.

\noindent \textbf{Organizing \fileformat{BAM} and \fileformat{SAM} files (Step~\ref{faffingbams}).}  The set of files containing mapped reads (from \tool{tophat2}, \emph{accepted\_hits.bam}) (typically) need to be transformed before they can be used with other downstream tools.  In particular, the \tool{samtools} command is used to prepare variations of the mapped reads.  Specifically, a sorted and indexed version of the \fileformat{BAM} file was created, which can be used in genome browsers such as \tool{IGV}; a sorted-by-name \fileformat{SAM} file was created, which is compatible with the feature counting software of \tool{htseq-count}.  Alternative feature counting tools (e.\,g., in \tool{Bioconductor}) may require different inputs.

\noindent \textbf{Design matrix.} For more complex designs (i.\,e., beyond two-group comparisons), users need to provide a design matrix that specifies the factors that are expected to affect expression levels.  As mentioned above, GLMs can be used to analyze arbitrarily complex experiments, and the design matrix is the means by which the experimental design is described mathematically, including both biological factors of interest and other factors not of direct interest, such as batch effects. For example, Section 4.5 of the \tool{edgeR} User's Guide (``RNA-Seq of pathogen inoculated Arabidopsis with batch effects'') or Section~4 of the \tool{DESeq} vignette (``Multi-factor designs'') present worked case studies with batch effects.
The design matrix is central for such complex differential expression analyses, and users may wish to consult with a linear modeling textbook\cite{Myers2000} or with a local statistician to make sure their design matrix is appropriately specified.

\noindent \textbf{Reproducible research.} So that other researchers (e.\,g., collaborators,
reviewers) can reproduce data analyses, we recommend that users keep a record of all commands (R and Unix)
and the software versions used in their analysis (for example, see Box~\ref{versionbox}). In practice, this is best achieved
by keeping the complete transcript of the computer commands interweaved with the textual
narrative in a single, executable document\cite{Gentleman2005}.	

\tool{R} provides many tools to facilitate the authoring of executable documents, including the \Rfunction{Sweave} function and the \tool{knitR} package. The \Rfunction{sessionInfo} function helps with documenting package versions and related information. A recent integration with \tool{Rstudio} is \url{rpubs.com}, which provides seamless integration of ``mark-down'' text with \tool{R} commands for easy web-based display. For language-independent authoring, a powerful tool is provided by \tool{Emacs org-mode}.


\section*{MATERIALS}

\subsection*{Equipment}

\noindent \emph{Operating system:} This protocol assumes users have a Unix-like
operating system, i.\,e., Linux or MacOS X, with a bash shell or similar.
All commands given here are meant to be run in a terminal
window. While it is possible to follow this protocol with a Microsoft
Windows machine (e.\,g., using Unix-like Cygwin; \url{http://www.cygwin.com/}), the additional steps
required are not discussed here.

\noindent \emph{Software:} Users will need the following software:

\begin{compactitem}

\item an aligner to map short reads to a genome that is able to deal
  with reads that straddle introns\cite{Fonseca2012}. 
  The aligner \tool{tophat2}\cite{Trapnell2009a,Trapnell2009} is
  illustrated here, but others, such as \tool{GSNAP}\cite{Wu2010},
  \tool{SpliceMap}\cite{Wang2010}, \tool{Subread}\cite{Shi2013} or \tool{STAR}\cite{Dobin2012}, among others, can be used.

\item optionally, a tool to visualize alignment files, such as the
  \tool{Integrated Genome Viewer (IGV)}\cite{Thorvaldsdottir2012}, or \tool{Savant}\cite{Fiume2010,Fiume2012}. \tool{IGV}
  is a Java tool with ``web start'' (downloadable from \url{http://www.broadinstitute.org/software/igv/download}), i.\,e., it can be started from
  a web browser and needs no explicit installation at the operating system level, provided a Java Runtime Environment is available.  
  
\item the \tool{R} statistical computing environment, downloadable from \url{http://www.r-project.org/}.

\item a number of \tool{Bioconductor}\cite{Gentleman2004} packages,
  specifically \tool{ShortRead}\cite{Morgan2009}, \tool{DESeq}\cite{Anders2010} and
  \tool{edgeR}\cite{Robinson2010a,McCarthy2012}, and possibly \tool{GenomicRanges},
  \tool{GenomicFeatures} and \tool{org.Dm.eg.db}, as well as their
  dependencies.

\item the \tool{samtools} program\cite{Li2009}, downloadable from \url{http://samtools.sourceforge.net/}, for manipulation of \fileformat{SAM} and \fileformat{BAM} formatted files.

\item the \tool{HTSeq} package, downloadable from \url{http://www-huber.embl.de/users/anders/HTSeq/doc/overview.html}, for counting of mapped reads.

\item optionally, if users wish to work with data from the Short
  Read Archive, the \tool{SRA Toolkit}, available from
  \url{http://www.ncbi.nlm.nih.gov/Traces/sra/sra.cgi?cmd=show&f=software&m=software&s=software}. 
  

\end{compactitem}

\noindent CRITICAL: For many of these software packages, new features and optimizations
are constantly developed and released, so it is highly recommended to use the most recent stable version as well as reading the (corresponding) documentation for the version used, since recommendations can change over time. The package
versions used in the production of this article are given in Box \ref{versionbox}.

\noindent \emph{Input file formats:} In general, the starting point is a collection of
\fileformat{FASTQ} files, the commonly used format for reads from Illumina 
sequencing machines.  The modifications necessary for mapping reads 
from other platforms are not discussed here.

\noindent \emph{Example data}: The data set published by Brooks et al.\cite{Brooks2011} is used 
here to demonstrate the workflow. This data set consists of seven
RNA-seq samples, each a cell culture of \emph{Drosophila melanogaster}
S2 cells. Three samples were treated with siRNA targeting the
splicing factor \emph{pasilla} (CG1844) (``Knockdown'') and four samples are
untreated (``Control''). Our aim is to identify genes that change in expression
between Knockdown and Control. 

\noindent Brooks et al.\cite{Brooks2011} have sequenced some of their libraries in single-end and others in paired-end
mode. This allows us to demonstrate two variants of the workflow: If we ignore the
differences in library type, the samples only differ by their experimental condition,
knockdown or control, and the analysis is a simple comparison between two sample groups.
We refer to this setting as an experiment with a \emph{simple design}. If we want to
account for library type as a blocking factor, our samples differ in more than one aspect,
i.e., we have a \emph{complex design}. To deal with the latter, we use \tool{edgeR} and \tool{DESeq}'s
functions to fit generalized linear models (GLMs).

\subsection*{Equipment setup}

\vspace{-3.5ex}
\subsubsection*{Install \tool{bowtie2}, \tool{tophat2} and \tool{samtools}}

\noindent Download and install \tool{samtools} from \url{http://samtools.sourceforge.net}.

\tool{bowtie2} and \tool{tophat2} have binary versions available for
Linux and Mac OS X platforms.  These can be downloaded from
\url{http://bowtie-bio.sourceforge.net/index.shtml} and
\url{http://tophat.cbcb.umd.edu}.  Consult the documentation on those
sites for further information if necessary.

\subsubsection*{Install \tool{R} and required \tool{Bioconductor} packages}

Download the \textbf{latest} release version of \tool{R} (at time of writing, \tool{R} version 3.0.0) from \url{http://cran.r-project.org} and install it. Consult the
\href{http://cran.r-project.org/doc/manuals/r-release/R-admin.pdf}{R Installation and Administration} manual if necessary.  A useful quick reference for R commands can be found at \url{http://cran.r-project.org/doc/contrib/Short-refcard.pdf}.

\noindent To install \tool{Bioconductor} packages, start \tool{R} by issuing
the command \texttt{R} in a terminal window and type:

\begin{Schunk}
\begin{Sinput}
> source( "http://www.bioconductor.org/biocLite.R" )
> biocLite("BiocUpgrade")
> biocLite( c("ShortRead","DESeq", "edgeR") )
\end{Sinput}
\end{Schunk}
This retrieves an automatic installation tool (\Rfunction{biocLite}) and installs the
version-matched packages. In addition, the installation tool
will automatically download and install all other packages that are prerequisite. 
Versions of \tool{Bioconductor} packages are matched to
versions of \tool{R}. Hence, to use current versions of \tool{Bioconductor} packages, it
is necessary to use a current version of \tool{R}.
Note that \tool{R} and \tool{Bioconductor}, at all times, maintain a stable \emph{release} version and
a \emph{development} version.  Unless a special need exists for a particular new
functionality, users should use the release version.


\subsubsection*{Download the example data}

CRITICAL: This step is only required if data originate from the Short Read Archive (SRA).

\noindent Brooks et al.\cite{Brooks2011} deposited their data in the Short Read Archive (SRA) of
the NCBI's Gene Expression Omnibus (\tool{GEO})\cite{Edgar2002} under accession number GSE18508 (\url{http://www.ncbi.nlm.nih.gov/geo/query/acc.cgi?acc=GSE18508}), and a subset of this data set is used here to illustrate the pipeline.  Specifically,  \fileformat{sra} files corresponding to the 4 ``Untreated'' (Control) and 3 ``CG8144\_RNAi'' (Knockdown) samples need to be downloaded.

\begin{figure}
\centering
\includegraphics[width=\textwidth]{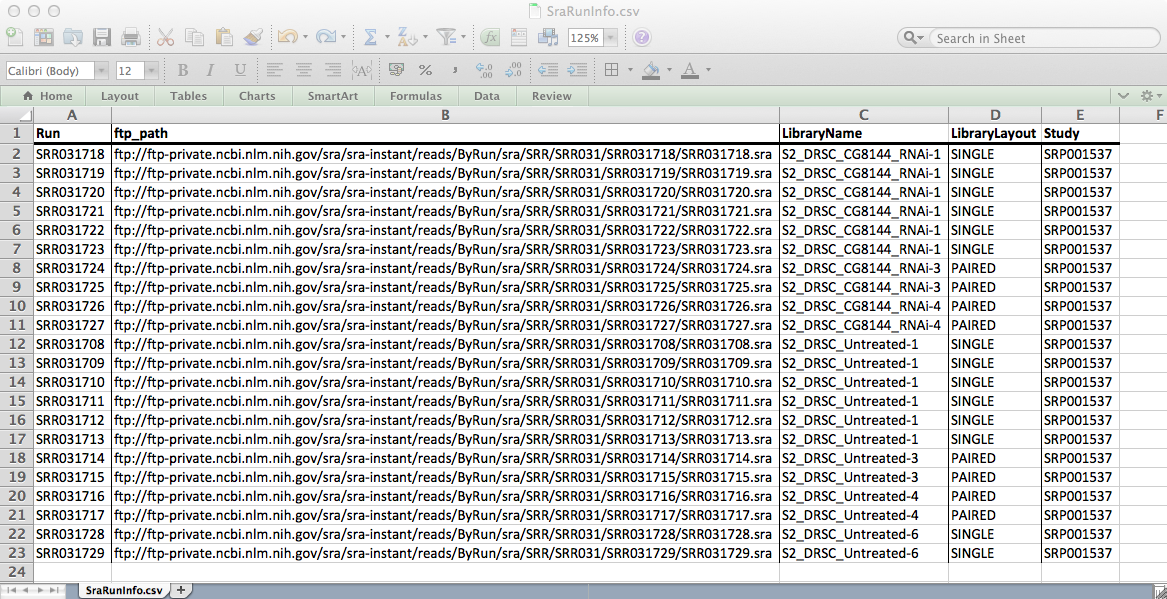}
\caption{Metadata available from Short Read Archive.}
\label{sra-meta}
\end{figure}

\noindent For downloading \tool{SRA} repository data, an automated process may be desirable. For example, from \url{http://www.ncbi.nlm.nih.gov/sra?term=SRP001537} (the entire experiment corresponding to GEO accession GSE18508), users can download a table of the metadata into a comma-separated tabular file ``SraRunInfo.csv'' (see Supplementary File 1, which contains an archive of various files used in this protocol).  To do this, click on ``Send to:'' (top right corner), select ``File'', select format ``RunInfo'' and click on ``Create File''.  Read this \fileformat{CSV} file ``SraRunInfo.csv'' into \tool{R}, and select the subset of samples that we are interested in (using \tool{R}'s string matching function \Rfunction{grep}), corresponding to the 22 \fileformat{SRA} files shown in Figure~\ref{sra-meta}  by:

\begin{Schunk}
\begin{Sinput}
> sri = read.csv("SraRunInfo.csv", stringsAsFactors=FALSE)
> keep = grep("CG8144|Untreated-",sri$LibraryName) 
> sri = sri[keep,] 
\end{Sinput}
\end{Schunk}

\noindent The following \tool{R} commands automate the download of the 22 \fileformat{SRA} files to the current working directory (the functions \Rfunction{getwd} and \Rfunction{setwd} can be used to retrieve and set the working directory, respectively):

\begin{Schunk}
\begin{Sinput}
> fs = basename(sri$download_path)
> for(i in 1:nrow(sri))
   download.file(sri$download_path[i], fs[i])
\end{Sinput}
\end{Schunk}

\noindent The \tool{R}-based download of files described above is just one way to capture several files in a semi-automatic fashion.  Users can alternatively use the batch tools \verb|wget| (Unix/Linux) or \verb|curl| (Mac OS X), or download using a web browser.
The (truncated) verbose output of the above \tool{R} download commands looks as follows:

\begin{Soutput}
trying URL 'ftp://ftp-private.ncbi.nlm.nih.gov/sra/sra-instant/reads/ByRun/sra/SRR/SRR031/SRR031714/SRR031714.sra'
ftp data connection made, file length 415554366 bytes
opened URL
=================================================
downloaded 396.3 Mb

trying URL 'ftp://ftp-private.ncbi.nlm.nih.gov/sra/sra-instant/reads/ByRun/sra/SRR/SRR031/SRR031715/SRR031715.sra'
ftp data connection made, file length 409390212 bytes
opened URL
================================================
downloaded 390.4 Mb
[... truncated ...]
\end{Soutput}

\subsubsection*{Convert SRA to \fileformat{FASTQ} format}

Typically, sequencing data from a sequencing facility will come in (compressed) \fileformat{FASTQ} format.
The SRA, however, uses its own, compressed, \fileformat{sra} format. To convert the example
data to \fileformat{FASTQ}, use the \tool{fastq-dump} command
from the \tool{SRA Toolkit} on each \fileformat{SRA} file. \tool{R} can be used to construct the 
required shell commands, starting from the ``SraRunInfo.csv'' metadata table, as follows:

\begin{Schunk}
\begin{Sinput}
> stopifnot( all(file.exists(fs)) )  # assure FTP download was successful
> for(f in fs) {
   cmd = paste("fastq-dump --split-3", f)
   cat(cmd,"\n")
   system(cmd) # invoke command
 }
\end{Sinput}
\end{Schunk}

\noindent CRITICAL: Using \tool{R}'s \Rfunction{system} command is just one possibility.  Users may choose to type the 22 \tool{fastq-dump} commands \textbf{manually} into the Unix shell rather than using \tool{R} to construct them.

\noindent CRITICAL: It is not absolutely necessary to use \Rfunction{cat} to print out the current 
command, but it serves the purpose of knowing what is currently running in the shell:

\begin{Soutput}
fastq-dump --split-3 SRR031714.sra 
Written 5327425 spots for SRR031714.sra
Written 5327425 spots total
fastq-dump --split-3 SRR031715.sra 
Written 5248396 spots for SRR031715.sra
Written 5248396 spots total
[... truncated ...]
\end{Soutput}

\noindent CRITICAL: Be sure to use the \verb|--split-3| option, which splits mate-pair reads into separate files.  After this command, single and paired-end data will produce one or two \fileformat{fastq} files, respectively.  For paired-end data, the file names will be suffixed \fileformat{\_1.fastq} and \fileformat{\_2.fastq}; otherwise, a single file with extension \fileformat{.fastq} will be produced.

\vspace{-3.5ex}
\subsubsection*{Download the reference genome} \label{dlref}

Download reference genome sequence for the organism under study in (compressed) \fileformat{FASTA} format.  Some useful resources, among others, include:

\begin{itemize}
\item the general Ensembl FTP server (\url{http://www.ensembl.org/info/data/ftp/index.html})
\item the Ensembl plants FTP server (\url{http://plants.ensembl.org/info/data/ftp/index.html})
\item the Ensembl metazoa FTP server (\url{http://metazoa.ensembl.org/info/data/ftp/index.html})
\item the UCSC current genomes FTP server (\url{ftp://hgdownload.cse.ucsc.edu/goldenPath/currentGenomes/})
\end{itemize}

\noindent For \tool{Ensembl}, choose the ``FASTA (DNA)'' link instead of ``FASTA (cDNA)'', since alignments to the genome, not the transcriptome, are desired.  For \emph{Drosphila melanogaster}, the file labeled ``toplevel'' combines all chromosomes. Do not use
the ``repeat-masked'' files (indicated by ``rm'' in the file name), since handling 
repeat regions should be left to the alignment algorithm.

\noindent The Drosophila reference genome can be downloaded from \tool{Ensembl} and uncompressed using the following Unix commands:

\begin{verbatim}
wget ftp://ftp.ensembl.org/pub/release-70/fasta/drosophila_melanogaster/\
dna/Drosophila_melanogaster.BDGP5.70.dna.toplevel.fa.gz

gunzip Drosophila_melanogaster.BDGP5.70.dna.toplevel.fa.gz
\end{verbatim}

\noindent For genomes provided by UCSC, users can select their genome of interest, proceed to the ``bigZips'' directory and download the ``chromFa.tar.gz''; as above, this could be done using the \texttt{wget} command.  Note that \tool{bowtie2}/\tool{tophat2} indices for many commonly used reference genomes can be downloaded directly from \url{http://tophat.cbcb.umd.edu/igenomes.html}.

\subsubsection*{Get gene model annotations}

Download a \fileformat{GTF} file with gene models for the organism of interest. For species covered by
\tool{Ensembl}, the Ensembl FTP site mentioned above contains links to such files.

\noindent The gene model annotation for \emph{Drosophila melanogaster} can be downloaded and uncompressed using:

\begin{verbatim}
wget ftp://ftp.ensembl.org/pub/release-70/gtf/drosophila_melanogaster/\
Drosophila_melanogaster.BDGP5.70.gtf.gz

gunzip Drosophila_melanogaster.BDGP5.70.gtf.gz
\end{verbatim}

\noindent CRITICAL: Make sure that the gene annotation uses the same coordinate system as the
reference \fileformat{FASTA} file. Here, both files use BDGP5 (i.\,e., release 5 of the assembly provided by the Berkeley Drosophila Genome Project), as is apparent from the file names. To be on the safe side here, we recommend to always download
the \fileformat{FASTA} reference sequence and the \fileformat{GTF} annotation
data from the same resource provider.

\noindent CRITICAL: As an alternative, the UCSC Table Browser (\url{http://genome.ucsc.edu/cgi-bin/hgTables}) can be used to generate \fileformat{GTF} files based on a selected annotation (e.\,g., RefSeq genes). 
However, at the time of writing \fileformat{GTF} files obtained from the UCSC Table Browser do not contain correct gene IDs, which causes
problems with downstream tools such as \tool{htseq-count}, unless corrected manually.

\subsubsection*{Build the reference index} \label{buildIndex}

Before reads can be aligned, the reference \fileformat{FASTA} files need to be preprocessed
into an \emph{index} that allows the aligner easy access. To build a \tool{bowtie2}-specific
index from the \fileformat{FASTA} file mentioned above, use the command:

\begin{verbatim}
bowtie2-build -f Drosophila_melanogaster.BDGP5.70.dna.toplevel.fa Dme1_BDGP5_70
\end{verbatim}


\noindent A set of \fileformat{BT2} files will be produced, with names starting with Dme1\_BDGP5\_70 as specfied above. This procedure needs to be run only once for each reference genome used.  As mentioned, pre-built indices for many commonly-used genomes are available from \url{http://tophat.cbcb.umd.edu/igenomes.html}.

\section*{PROCEDURE}
\setcounter{step}{0}

\subsection{Assess sequence quality control with \tool{ShortRead}}
\Timing{$\sim$2 hours}


\Step{At the \tool{R} prompt, type the commands (you may first need to use \Rfunction{setwd} to set the working directory to where the \fileformat{FASTQ} files are situated):}
\label{qc}

\begin{Schunk}
\begin{Sinput}
> library("ShortRead")
> fqQC = qa(dirPath=".", pattern=".fastq$", type="fastq")
> report(fqQC, type="html", dest="fastqQAreport")
\end{Sinput}
\end{Schunk}

\Step{Use a web browser to inspect the generated \fileformat{HTML} file (here, stored in the ``fastqQAreport'' directory) with the quality-assessment report.}


\subsection{Collect metadata of experimental design} \ \\


\Step{Create a table of metadata called \Robject{samples} (see Constructing metadata table).  This step needs
to be adapted for each data set, and many users may find a spreadsheet
program useful for this step, from which data can be
imported into the table \Robject{samples} by the \Rfunction{read.csv}
function.  For our example data, we chose to construct the
\Robject{samples} table programmatically from the table of
\fileformat{SRA} files.}\label{meta-start}

\Step{Collapse the initial table \Robject{sri} to one row per sample:}

\begin{Schunk}
\begin{Sinput}
> sri$LibraryName = gsub("S2_DRSC_","",sri$LibraryName) # trim label
> samples = unique(sri[,c("LibraryName","LibraryLayout")])
> for(i in seq_len(nrow(samples))) {
   rw = (sri$LibraryName==samples$LibraryName[i])
   if(samples$LibraryLayout[i]=="PAIRED") {
     samples$fastq1[i] = paste0(sri$Run[rw],"_1.fastq",collapse=",")
     samples$fastq2[i] = paste0(sri$Run[rw],"_2.fastq",collapse=",")
   } else {
     samples$fastq1[i] = paste0(sri$Run[rw],".fastq",collapse=",")
     samples$fastq2[i] = ""
   }
 }
\end{Sinput}
\end{Schunk}



\Step{Add important or descriptive columns to the metadata table
(here, experimental groupings are set based on the ``LibraryName''
column, and a label is created for plotting):} 

\begin{Schunk}
\begin{Sinput}
> samples$condition = "CTL"
> samples$condition[grep("RNAi",samples$LibraryName)] = "KD"
> samples$shortname = paste( substr(samples$condition,1,2), 
                             substr(samples$LibraryLayout,1,2), 
                             seq_len(nrow(samples)), sep=".")
\end{Sinput}
\end{Schunk}

\Step{Since the downstream statistical analysis of differential expression relies on this table, carefully inspect (and correct, if necessary) the metadata table.  In particular, verify that there exists one row per sample, that all columns of information are populated and the files names, labels and experimental conditions are correct.} \label{meta-end}

{\footnotesize  
\begin{Schunk}
\begin{Sinput}
> samples
\end{Sinput}
\end{Schunk}
\begin{Schunk}
\begin{Soutput}
    LibraryName LibraryLayout                fastq1                fastq2 condition shortname
1   Untreated-3        PAIRED SRR031714_1.fastq,... SRR031714_2.fastq,...       CTL   CT.PA.1
2   Untreated-4        PAIRED SRR031716_1.fastq,... SRR031716_2.fastq,...       CTL   CT.PA.2
3 CG8144_RNAi-3        PAIRED SRR031724_1.fastq,... SRR031724_2.fastq,...        KD   KD.PA.3
4 CG8144_RNAi-4        PAIRED SRR031726_1.fastq,... SRR031726_2.fastq,...        KD   KD.PA.4
5   Untreated-1        SINGLE   SRR031708.fastq,...                             CTL   CT.SI.5
6 CG8144_RNAi-1        SINGLE   SRR031718.fastq,...                              KD   KD.SI.6
7   Untreated-6        SINGLE   SRR031728.fastq,...                             CTL   CT.SI.7
\end{Soutput}
\end{Schunk}
}



\subsection{Align the reads (using \tool{tophat2}) to reference genome}
\Timing{$\sim$45 minutes per sample}




\Step{Using \tool{R} string manipulation, construct the Unix commands to call \tool{tophat2}. Given the metadata table \Robject{samples}, it is convenient to use \tool{R} to create the list of shell commands, as follows:}
\label{mapping-start}

\begin{Schunk}
\begin{Sinput}
> gf = "Drosophila_melanogaster.BDGP5.70.gtf"
> bowind = "Dme1_BDGP5_70"
> cmd = with(samples, 
   paste("tophat -G", gf, "-p 5 -o", LibraryName, bowind, 
         fastq1, fastq2))
\end{Sinput}
\end{Schunk}
\begin{Schunk}
\begin{Sinput}
> cmd
\end{Sinput}
\end{Schunk}
\begin{Schunk}
\begin{Soutput}
tophat -G Drosophila_melanogaster.BDGP5.70.gtf -p 5 -o Untreated-3 Dme1_BDGP5_70 \
SRR031714_1.fastq,SRR031715_1.fastq SRR031714_2.fastq,SRR031715_2.fastq 

tophat -G Drosophila_melanogaster.BDGP5.70.gtf -p 5 -o Untreated-4 Dme1_BDGP5_70 \
SRR031716_1.fastq,SRR031717_1.fastq SRR031716_2.fastq,SRR031717_2.fastq 

tophat -G Drosophila_melanogaster.BDGP5.70.gtf -p 5 -o CG8144_RNAi-3 Dme1_BDGP5_70 \
SRR031724_1.fastq,SRR031725_1.fastq SRR031724_2.fastq,SRR031725_2.fastq 

tophat -G Drosophila_melanogaster.BDGP5.70.gtf -p 5 -o CG8144_RNAi-4 Dme1_BDGP5_70 \
SRR031726_1.fastq,SRR031727_1.fastq SRR031726_2.fastq,SRR031727_2.fastq 

tophat -G Drosophila_melanogaster.BDGP5.70.gtf -p 5 -o Untreated-1 Dme1_BDGP5_70 \
SRR031708.fastq,SRR031709.fastq,SRR031710.fastq,SRR031711.fastq,SRR031712.fastq,SRR031713.fastq 

tophat -G Drosophila_melanogaster.BDGP5.70.gtf -p 5 -o CG8144_RNAi-1 Dme1_BDGP5_70 \
SRR031718.fastq,SRR031719.fastq,SRR031720.fastq,SRR031721.fastq,SRR031722.fastq,SRR031723.fastq 

tophat -G Drosophila_melanogaster.BDGP5.70.gtf -p 5 -o Untreated-6 Dme1_BDGP5_70 \
SRR031728.fastq,SRR031729.fastq 
\end{Soutput}
\end{Schunk}

\noindent CRITICAL: In the call to \tool{tophat2}, the option \texttt{-G} points \tool{tophat2} to a \fileformat{GTF} file of annotation to facilitate mapping reads across exon-exon junctions (some of which can be found \emph{de novo}), \texttt{-o} specifies the output directory, \texttt{-p} specifies the number of threads to use (this may affect run times and can vary depending on the resources available).  Other parameters can be specified here, as needed; see the appropriate documentation for the tool and version you are using.  The first argument, \texttt{Dmel\_BDGP5\_70} is the name of the index (built in advance), and the second argument is a list of all \fileformat{FASTQ} files with reads for the sample. Note that the \fileformat{FASTQ} files are concatenated with commas, \emph{without} spaces.  For experiments with paired-end reads, pairs of \fileformat{fastq} files are given as separate arguments and the order in both arguments must match.  

\Step{Run these commands (i.e. copy-and-paste) in a Unix terminal.}
\label{mapping-end}

\noindent CRITICAL: Many similar possibilities exist for this step.  Users can use the \tool{R} function \Rfunction{system} to execute these commands direct from \tool{R}, cut-and-paste the commands into a separate Unix shell or store the list of commands in a text file and use the Unix \texttt{source} command.  In addition, users could construct the unix commands independent of \tool{R}.

\subsection{Organize, sort and index the \fileformat{BAM} files and create \fileformat{SAM} files}
\Timing{$\sim$1 hour}

\Step{Organize the \fileformat{BAM} files into a single directory, sort and index them and create SAM files, by running the following R-generated commands:}
\label{faffingbams}

\begin{Schunk}
\begin{Sinput}
> for(i in seq_len(nrow(samples))) {
   lib = samples$LibraryName[i]
   ob = file.path(lib, "accepted_hits.bam")
 
   # sort by name, convert to SAM for htseq-count
   cat(paste0("samtools sort -n ",ob," ",lib,"_sn"),"\n")
   cat(paste0("samtools view -o ",lib,"_sn.sam ",lib,"_sn.bam"),"\n")
 
   # sort by position and index for IGV
   cat(paste0("samtools sort ",ob," ",lib,"_s"),"\n")
   cat(paste0("samtools index ",lib,"_s.bam"),"\n\n")
 }
\end{Sinput}
\begin{Soutput}
samtools sort -n Untreated-3/accepted_hits.bam Untreated-3_sn 
samtools view -o Untreated-3_sn.sam Untreated-3_sn.bam 
samtools sort Untreated-3/accepted_hits.bam Untreated-3_s 
samtools index Untreated-3_s.bam 

samtools sort -n Untreated-4/accepted_hits.bam Untreated-4_sn 
samtools view -o Untreated-4_sn.sam Untreated-4_sn.bam 
samtools sort Untreated-4/accepted_hits.bam Untreated-4_s 
samtools index Untreated-4_s.bam 

samtools sort -n CG8144_RNAi-3/accepted_hits.bam CG8144_RNAi-3_sn 
samtools view -o CG8144_RNAi-3_sn.sam CG8144_RNAi-3_sn.bam 
samtools sort CG8144_RNAi-3/accepted_hits.bam CG8144_RNAi-3_s 
samtools index CG8144_RNAi-3_s.bam 

samtools sort -n CG8144_RNAi-4/accepted_hits.bam CG8144_RNAi-4_sn 
samtools view -o CG8144_RNAi-4_sn.sam CG8144_RNAi-4_sn.bam 
samtools sort CG8144_RNAi-4/accepted_hits.bam CG8144_RNAi-4_s 
samtools index CG8144_RNAi-4_s.bam 

samtools sort -n Untreated-1/accepted_hits.bam Untreated-1_sn 
samtools view -o Untreated-1_sn.sam Untreated-1_sn.bam 
samtools sort Untreated-1/accepted_hits.bam Untreated-1_s 
samtools index Untreated-1_s.bam 

samtools sort -n CG8144_RNAi-1/accepted_hits.bam CG8144_RNAi-1_sn 
samtools view -o CG8144_RNAi-1_sn.sam CG8144_RNAi-1_sn.bam 
samtools sort CG8144_RNAi-1/accepted_hits.bam CG8144_RNAi-1_s 
samtools index CG8144_RNAi-1_s.bam 

samtools sort -n Untreated-6/accepted_hits.bam Untreated-6_sn 
samtools view -o Untreated-6_sn.sam Untreated-6_sn.bam 
samtools sort Untreated-6/accepted_hits.bam Untreated-6_s 
samtools index Untreated-6_s.bam 
\end{Soutput}
\end{Schunk}

\noindent CRITICAL: Users should be conscious of the disk space that may get used in these operations.  In the command above, sorted-by-name \fileformat{SAM} and \fileformat{BAM} files (for \tool{htseq-count}), as well as a sorted-by-chromosome-position \fileformat{BAM} file (for \tool{IGV}) are created for each original \emph{accepted\_hits.bam} file.  User may wish to delete (some of) these files after the steps below.

\subsection{Inspect alignments with IGV}  \label{igv}

\ \\

\Step{Start \tool{IGV}, select the correct genome (here,
\emph{D.~melanogaster (dm3)}) and load the \fileformat{BAM} files
(with \_s in the filename) as well as the \fileformat{GTF} file.} \label{igv-start}

\begin{figure}
\centering
\includegraphics[width=\textwidth]{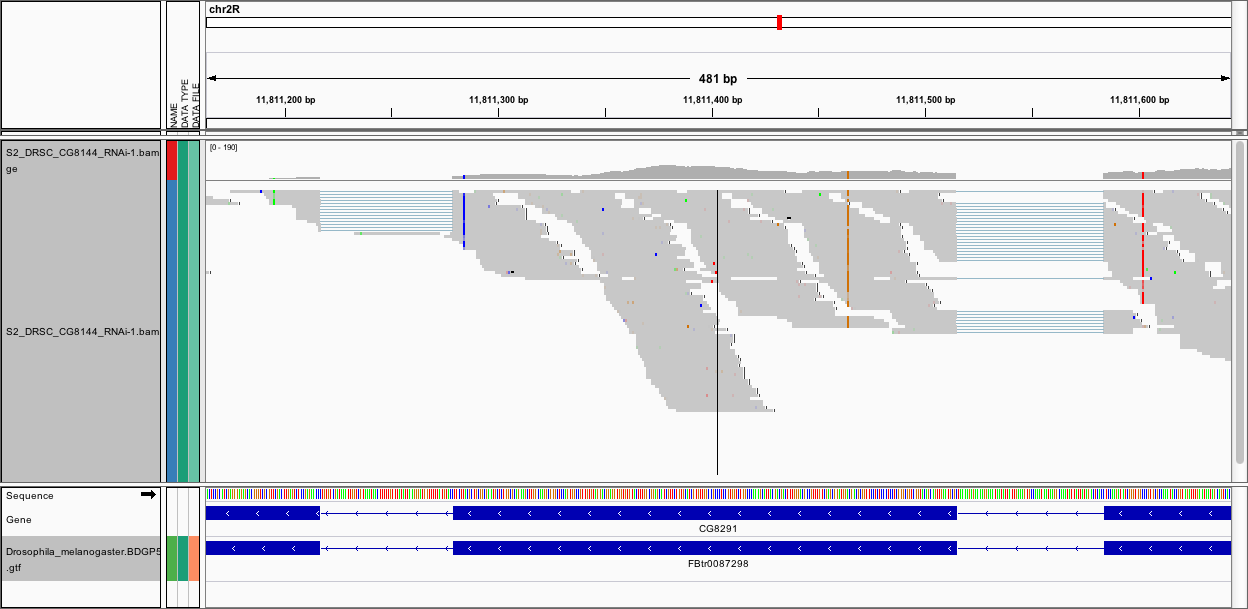}
\caption{Screenshot of reads aligning across exon junctions.}
\label{igv-plot}
\end{figure}

\Step{Zoom in on an expressed transcript until individual reads are shown and check
whether the reads align at and across exon-exon junctions, as expected
given the annotation (See example in Figure~\ref{igv-plot}).}

\Step{If any positive and negative controls are known for the system under study (e.g. known differential expression), 
direct the IGV browser to these regions to confirm that the relative read density is different according to expectation.} \label{igv-end}

\subsection{Count reads using \tool{htseq-count}}
\Timing{$\sim$3 hours}


\Step{\noindent Add the names of the \fileformat{count} files to the metadata table and call \tool{HTSeq} from the following R-generated Unix commands:}
\label{counting}

\begin{Schunk}
\begin{Sinput}
> samples$countf = paste(samples$LibraryName, "count", sep=".")                            
\end{Sinput}
\end{Schunk}

\begin{Schunk}
\begin{Sinput}
> gf = "Drosophila_melanogaster.BDGP5.70.gtf"
> cmd = paste0("htseq-count -s no -a 10 ", samples$LibraryName, "_sn.sam ",
               gf," > ", samples$countf)
\end{Sinput}
\end{Schunk}
\begin{Schunk}
\begin{Sinput}
> cmd
\end{Sinput}
\end{Schunk}
\begin{Schunk}
\begin{Soutput}
htseq-count -s no -a 10 Untreated-3_sn.sam \
Drosophila_melanogaster.BDGP5.70.gtf > Untreated-3.count 

htseq-count -s no -a 10 Untreated-4_sn.sam \
Drosophila_melanogaster.BDGP5.70.gtf > Untreated-4.count 

htseq-count -s no -a 10 CG8144_RNAi-3_sn.sam \
Drosophila_melanogaster.BDGP5.70.gtf > CG8144_RNAi-3.count 

htseq-count -s no -a 10 CG8144_RNAi-4_sn.sam \
Drosophila_melanogaster.BDGP5.70.gtf > CG8144_RNAi-4.count 

htseq-count -s no -a 10 Untreated-1_sn.sam \
Drosophila_melanogaster.BDGP5.70.gtf > Untreated-1.count 

htseq-count -s no -a 10 CG8144_RNAi-1_sn.sam \
Drosophila_melanogaster.BDGP5.70.gtf > CG8144_RNAi-1.count 

htseq-count -s no -a 10 Untreated-6_sn.sam \
Drosophila_melanogaster.BDGP5.70.gtf > Untreated-6.count 
\end{Soutput}
\end{Schunk}

\noindent CRITICAL: The option \texttt{-s} signifies that the data is not from a stranded protocol (this may vary by experiment) and the \texttt{-a} option specifies a minimum score for the alignment quality.

\Step{For differential expression analysis with \tool{edgeR}, follow option A for simple designs and option B for complex designs; for differential expression analysis with \tool{DESeq}, follow option C for simple designs and option D for complex designs.}
\label{DEanalysis}

 

\subsection{A. \tool{edgeR} - simple design}



\begin{figure}
\centering
\includegraphics[width=\textwidth]{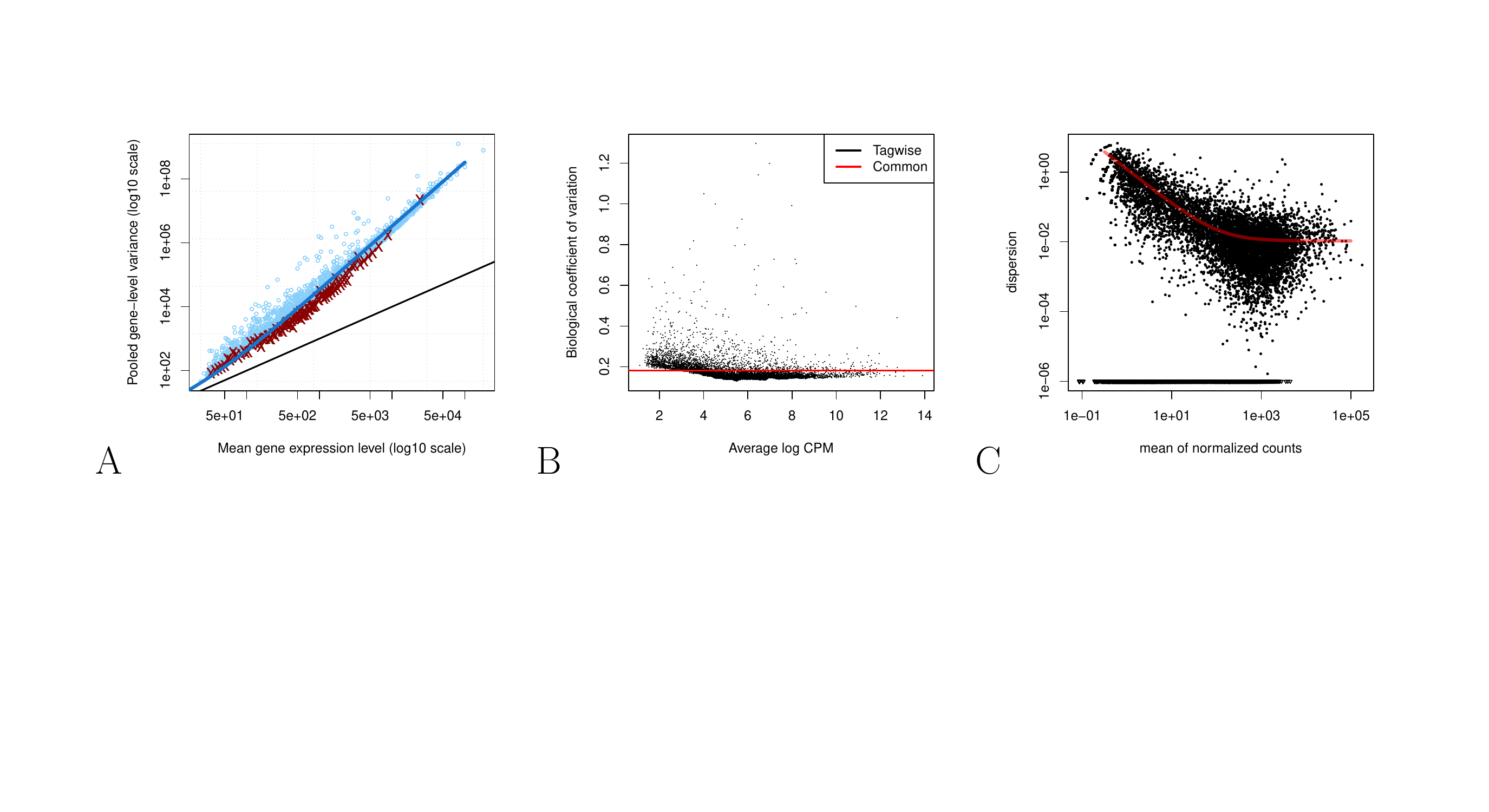}
\caption{Plots of mean-variance relationship and dispersion.  A. \tool{edgeR}'s \Rfunction{plotMeanVar} can be used for exploring the mean-variance relationship; each dot represents the estimated mean and variance for each gene, with binned variances as well as the trended common dispersion overlaid.  B. \tool{edgeR}'s \Rfunction{plotBCV} illustrates the relationship of biological coefficient of variation versus the mean.  C. \tool{DESeq}'s \Rfunction{plotDispEsts} shows the fit of dispersion versus mean.}
\label{fig_plotMeanVar_plotDispEsts}
\end{figure}

\begin{enumerate}[i)]

\item Load the \tool{edgeR} package and use the utility function, \Rfunction{readDGE}, to read in the \fileformat{COUNT} files created from \tool{htseq-count}: \label{load_edgeR} \label{edgeR-common-start}

\begin{Schunk}
\begin{Sinput}
> library("edgeR")
> counts = readDGE(samples$countf)$counts
\end{Sinput}
\end{Schunk}

%
\item Filter lowly expressed and non-informative (e.\,g., non-aligned) features using a command like:

\begin{Schunk}
\begin{Sinput}
> noint = rownames(counts) %in%
            c("no_feature","ambiguous","too_low_aQual",
              "not_aligned","alignment_not_unique")
> cpms = cpm(counts)
> keep = rowSums(cpms>1)>=3 & !noint
> counts = counts[keep,]
\end{Sinput}
\end{Schunk}

\noindent CRITICAL: In \tool{edgeR}, it is recommended to remove features without at least 1 read per
million in $n$ of the samples, where $n$ is the size of the smallest
group of replicates (here, $n=3$ for the Knockdown group).

\item Visualize and inspect the count table using:

\begin{Schunk}
\begin{Sinput}
> colnames(counts) = samples$shortname
> head( counts[,order(samples$condition)], 5 )
\end{Sinput}
\begin{Soutput}
            CT.PA.1 CT.PA.2 CT.SI.5 CT.SI.7 KD.PA.3 KD.PA.4 KD.SI.6
FBgn0000008      76      71     137      82      87      68     115
FBgn0000017    3498    3087    7014    3926    3029    3264    4322
FBgn0000018     240     306     613     485     288     307     528
FBgn0000032     611     672    1479    1351     694     757    1361
FBgn0000042   40048   49144   97565   99372   70574   72850   95760
\end{Soutput}
\end{Schunk}

\item Create a \Rclass{DGEList} object (\tool{edgeR}'s container for RNA-seq count data), as follows:

\begin{Schunk}
\begin{Sinput}
> d = DGEList(counts=counts, group=samples$condition)
\end{Sinput}
\end{Schunk}

\item Estimate normalization factors using:

\begin{Schunk}
\begin{Sinput}
> d = calcNormFactors(d)
> d$samples
\end{Sinput}
\end{Schunk}

\item Inspect the relationships between samples using a multidimensional scaling plot, as shown in Figure~\ref{fig_MDS_PCA}A:
\label{sampleRelations-edgeR} \label{edgeR-common-end}

\begin{Schunk}
\begin{Sinput}
> plotMDS(d, labels=samples$shortname, 
         col=c("darkgreen","blue")[factor(samples$condition)])
\end{Sinput}
\end{Schunk}








\item Estimate tagwise dispersion (simple design) using:

\begin{Schunk}
\begin{Sinput}
> d = estimateCommonDisp(d)
> d = estimateTagwiseDisp(d)
\end{Sinput}
\end{Schunk}

\item Create a visual representation of the mean-variance relationship using the \Rfunction{plotMeanVar} (shown in Figure~\ref{fig_plotMeanVar_plotDispEsts}A) and \Rfunction{plotBCV} (Figure~\ref{fig_plotMeanVar_plotDispEsts}B) functions, as follows:

\begin{Schunk}
\begin{Sinput}
> plotMeanVar(d, show.tagwise.vars=TRUE, NBline=TRUE)
\end{Sinput}
\end{Schunk}

\begin{Schunk}
\begin{Sinput}
> plotBCV(d)
\end{Sinput}
\end{Schunk}

\item Test for differential expression (``classic'' \tool{edgeR}), as follows:


\begin{Schunk}
\begin{Sinput}
> de = exactTest(d, pair=c("CTL","KD"))
\end{Sinput}
\end{Schunk}

\item Follow Step \ref{DEanalysis} B \ref{edgeR-common2-start})-\ref{edgeR-common2-end}).

\end{enumerate}




\subsection{B. \tool{edgeR} - complex design}

\begin{figure}
\centering
\includegraphics[width=\textwidth]{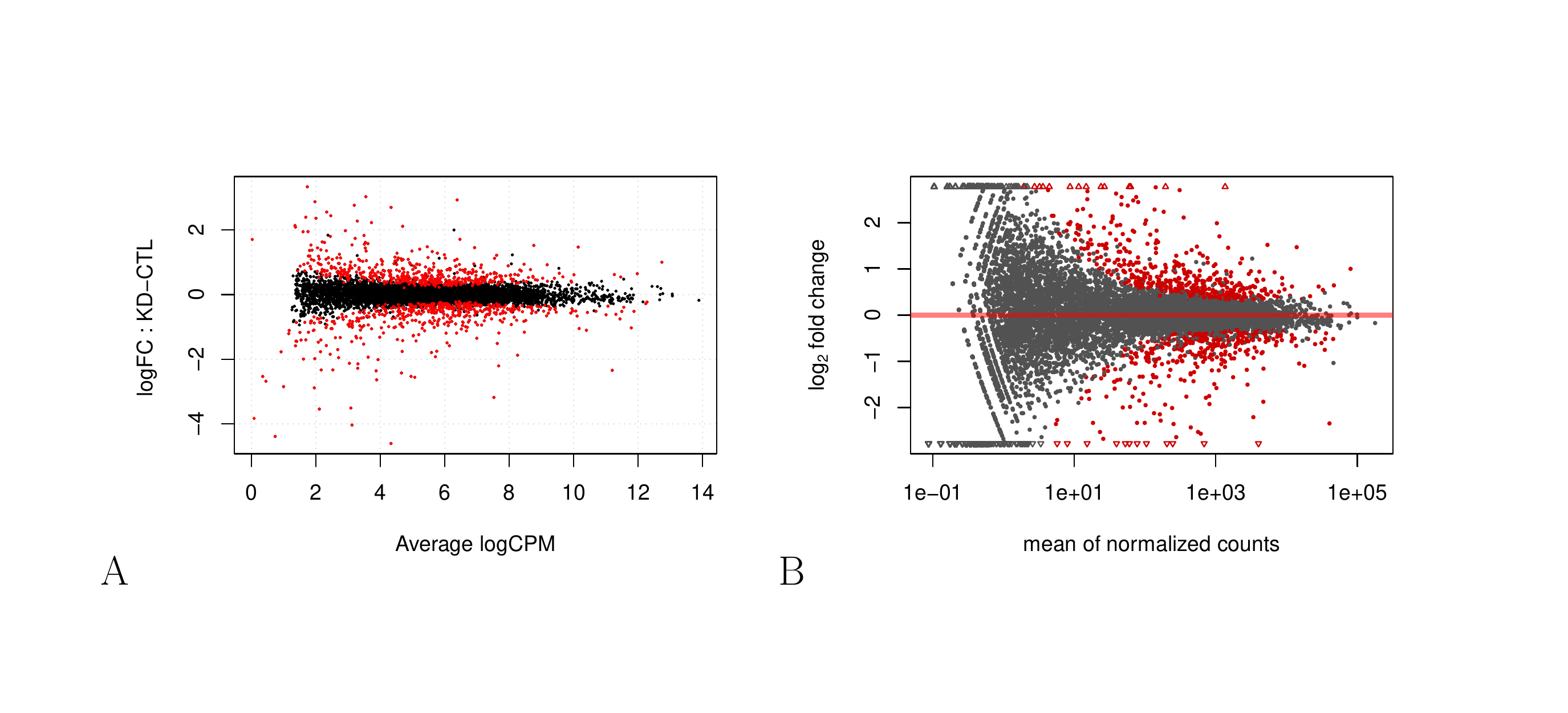}
\caption{M (``minus'') versus A (``add'') plots for RNA-seq data.  A. \tool{edgeR}'s \Rfunction{plotSmear} function plots the log-fold change (i.\,e., the log ratio of normalized expression levels between two experimental conditions) against the log-counts-per-million.  B. Similarly, \tool{DESeq}'s \Rfunction{plotMA} displays differential expression (log-fold-changes) versus expression strength (log-average-read-count).}
\label{fig_plotSmear_plotMA}
\end{figure}

\begin{enumerate}[i)]
\item Follow Step \ref{DEanalysis} A \ref{edgeR-common-start})-\ref{edgeR-common-end}).

\item Create a design matrix to specify the factors that are expected to affect expression levels:

\begin{Schunk}
\begin{Sinput}
> design = model.matrix( ~ LibraryLayout + condition, samples)
> design
\end{Sinput}
\begin{Soutput}
  (Intercept) LibraryLayoutSINGLE conditionKD
1           1                   0           0
2           1                   0           0
3           1                   0           1
4           1                   0           1
5           1                   1           0
6           1                   1           1
7           1                   1           0
attr(,"assign")
[1] 0 1 2
attr(,"contrasts")
attr(,"contrasts")$LibraryLayout
[1] "contr.treatment"

attr(,"contrasts")$condition
[1] "contr.treatment"
\end{Soutput}
\end{Schunk}

\item Estimate dispersion values, relative to the design matrix, using the Cox-Reid (CR) adjusted likelihood\cite{Cox1987,McCarthy2012}, as follows:

\begin{Schunk}
\begin{Sinput}
> d2 = estimateGLMTrendedDisp(d, design)
> d2 = estimateGLMTagwiseDisp(d2, design)
\end{Sinput}
\end{Schunk}


\item Given the design matrix and dispersion estimates, fit a GLM to each feature:

\begin{Schunk}
\begin{Sinput}
> f = glmFit(d2, design)
\end{Sinput}
\end{Schunk}

\item Perform a likelihood ratio test, specifying the difference of interest (here, Knockdown versus Control, which corresponds to the 3$^{rd}$ column of the above design matrix):

\begin{Schunk}
\begin{Sinput}
> de = glmLRT(f, coef=3)
\end{Sinput}
\end{Schunk}



\subsection{Inspect the results in graphical and tabular format} \label{inspectResults}

\item Use the \Rfunction{topTags} function to present a tabular summary of the differential expression statistics (Note: \Rfunction{topTags} operates on the output of \Rfunction{exactTest} or \Rfunction{glmLRT}, while only the latter is shown here): \label{edgeR-common2-start}

\begin{Schunk}
\begin{Sinput}
> tt = topTags(de, n=nrow(d))
> head(tt$table)
\end{Sinput}
\begin{Soutput}
            logFC logCPM  LR    PValue       FDR
FBgn0039155 -4.61   5.87 902 3.96e-198 2.85e-194
FBgn0025111  2.87   6.86 641 2.17e-141 7.81e-138
FBgn0039827 -4.05   4.40 457 2.11e-101  5.07e-98
FBgn0035085 -2.58   5.59 408  9.31e-91  1.68e-87
FBgn0000071  2.65   4.73 365  2.46e-81  3.54e-78
FBgn0003360 -3.12   8.42 359  3.62e-80  4.34e-77
\end{Soutput}
\end{Schunk}

\item Inspect the depth-adjusted reads per million for some of the top differentially expressed genes:

\begin{Schunk}
\begin{Sinput}
> nc = cpm(d, normalized.lib.sizes=TRUE)
> rn = rownames(tt$table)
> head(nc[rn,order(samples$condition)],5)
\end{Sinput}
\begin{Soutput}
            CT.PA.1 CT.PA.2 CT.SI.5 CT.SI.7 KD.PA.3 KD.PA.4 KD.SI.6
FBgn0039155   91.07    98.0  100.75  106.78    3.73    4.96    3.52
FBgn0025111   34.24    31.6   26.64   28.46  247.43  254.28  188.39
FBgn0039827   39.40    36.7   30.09   34.47    1.66    2.77    2.01
FBgn0035085   78.06    81.4   63.59   74.08   13.49   14.13   10.99
FBgn0000071    9.08     9.2    7.48    5.85   52.08   55.93   45.65
\end{Soutput}
\end{Schunk}

\item Create a graphical summary, such as an M (log-fold-change) versus A 
(log-average-expression) plot\cite{Dudoit2002}, here showing the genes selected as differentially expressed (with a 5\% false discovery rate; see Figure~\ref{fig_plotSmear_plotMA}A):

\begin{Schunk}
\begin{Sinput}
> deg = rn[tt$table$FDR < .05]
> plotSmear(d, de.tags=deg)
\end{Sinput}
\end{Schunk}

\subsection{Create persistent storage of results} 

\item Save the result table as a \fileformat{CSV} (comma-separated values) file (alternative formats are possible) as follows:
\label{persistent_storage_edgeR} \label{edgeR-common2-end}

\begin{Schunk}
\begin{Sinput}
> write.csv(tt$table, file="toptags_edgeR.csv")
\end{Sinput}
\end{Schunk}

\end{enumerate}






\subsection{C. \tool{DESeq} - simple design}

\begin{figure}
\centering
\includegraphics[width=.49\textwidth]{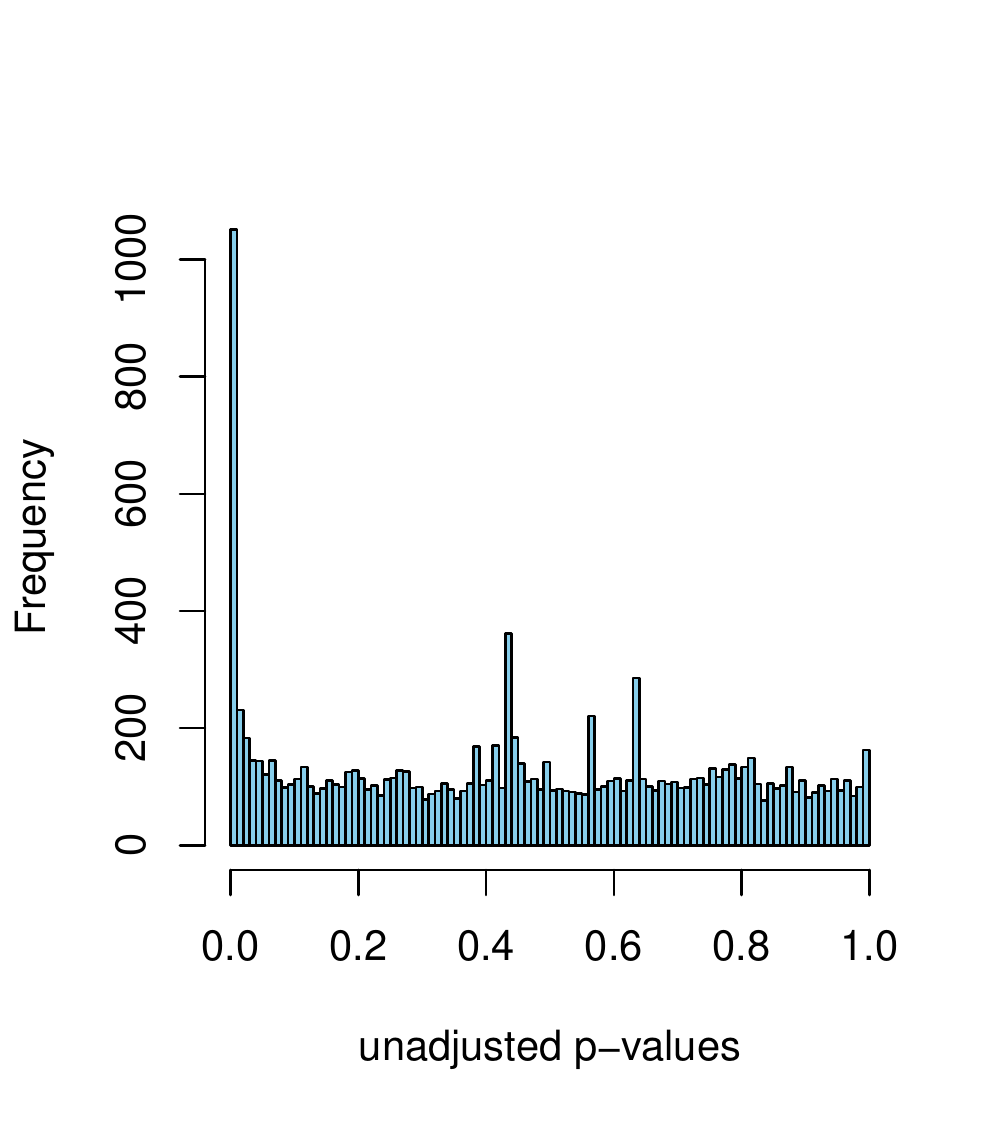}
\caption{Histogram of $P$-values from gene-by-gene statistical tests.}
\label{fighistp}
\end{figure}

\begin{enumerate}[i)]





\item Create a \Rclass{data.frame} with the required metadata, i.\,e., the names of the count files and experimental conditions.
Here, we derive it from the \Robject{samples} table created earlier.  \label{DESeq-common-start}
\begin{Schunk}
\begin{Sinput}
> samplesDESeq = with(samples, data.frame(
   shortname     = I(shortname),
   countf        = I(countf),
   condition     = condition,
   LibraryLayout = LibraryLayout))
\end{Sinput}
\end{Schunk}

\item Load the \tool{DESeq} package and create a \Rclass{CountDataSet} object
(\tool{DESeq}'s container for RNA-seq data) from the count tables and corresponding
metadata:  \label{load_deseq}

\begin{Schunk}
\begin{Sinput}
> library("DESeq")
> cds = newCountDataSetFromHTSeqCount(samplesDESeq)
\end{Sinput}
\end{Schunk}


\item Estimate normalization factors using:

\begin{Schunk}
\begin{Sinput}
> cds = estimateSizeFactors(cds)
\end{Sinput}
\end{Schunk}

\item Inspect the size factors using:

\begin{Schunk}
\begin{Sinput}
> sizeFactors(cds)
\end{Sinput}
\begin{Soutput}
CT.PA.1 CT.PA.2 KD.PA.3 KD.PA.4 CT.SI.5 KD.SI.6 CT.SI.7 
  0.699   0.811   0.822   0.894   1.643   1.372   1.104 
\end{Soutput}
\end{Schunk}


\item To inspect sample relationships, invoke a variance stabilizing transformation and 
inspect a principal component analysis (PCA) plot (shown in Figure~\ref{fig_MDS_PCA}B):
\label{sampleRelations-deseq} \label{DESeq-common-end}

\begin{Schunk}
\begin{Sinput}
> cdsB = estimateDispersions(cds, method="blind")
> vsd = varianceStabilizingTransformation(cdsB)
> p = plotPCA(vsd, intgroup=c("condition","LibraryLayout"))
\end{Sinput}
\end{Schunk}



\item Use \Rfunction{estimateDispersions} to calculate dispersion values:

\begin{Schunk}
\begin{Sinput}
> cds = estimateDispersions(cds)
\end{Sinput}
\end{Schunk}

\item Inspect the estimated dispersions using the \Rfunction{plotDispEsts} function (shown in Figure~\ref{fig_plotMeanVar_plotDispEsts}C), as follows:

\begin{Schunk}
\begin{Sinput}
> plotDispEsts(cds)
\end{Sinput}
\end{Schunk}


\item Perform the test for differential expression, using \Rfunction{nbinomTest}, as follows:

\begin{Schunk}
\begin{Sinput}
> res = nbinomTest(cds,"CTL","KD")
\end{Sinput}
\end{Schunk}

\item Given the table of differential expression results, use \Rfunction{plotMA} to display differential expression (log-fold-changes) versus expression strength (log-average-read-count), as follows (see Figure~\ref{fig_plotSmear_plotMA}B):

\begin{Schunk}
\begin{Sinput}
> plotMA(res)
\end{Sinput}
\end{Schunk}


\item Inspect the result tables of significantly up- and down-regulated, at 10\% FDR, using:

\begin{Schunk}
\begin{Sinput}
> resSig = res[which(res$padj < 0.1),]
> head( resSig[ order(resSig$log2FoldChange, decreasing=TRUE), ] )
\end{Sinput}
\begin{Soutput}
               id baseMean baseMeanA baseMeanB foldChange log2FoldChange     pval     padj
1515  FBgn0013696     1.46     0.000      3.40        Inf            Inf 4.32e-03 6.86e-02
13260 FBgn0085822     1.93     0.152      4.29       28.2           4.82 4.54e-03 7.11e-02
13265 FBgn0085827     8.70     0.913     19.08       20.9           4.39 1.02e-09 9.57e-08
15470 FBgn0264344     3.59     0.531      7.68       14.5           3.86 4.55e-04 1.10e-02
8153  FBgn0037191     4.43     0.715      9.39       13.1           3.71 5.35e-05 1.78e-03
1507  FBgn0013688    23.82     4.230     49.95       11.8           3.56 3.91e-21 1.38e-18
\end{Soutput}
\begin{Sinput}
> head( resSig[ order(resSig$log2FoldChange, decreasing=FALSE), ] )
\end{Sinput}
\begin{Soutput}
               id baseMean baseMeanA baseMeanB foldChange log2FoldChange      pval      padj
13045 FBgn0085359     60.0     102.2      3.78     0.0370          -4.76  7.65e-30  4.88e-27
9499  FBgn0039155    684.1    1161.5     47.59     0.0410          -4.61 3.05e-152 3.88e-148
2226  FBgn0024288     52.6      88.9      4.25     0.0478          -4.39  2.95e-32  2.09e-29
9967  FBgn0039827    246.2     412.0     25.08     0.0609          -4.04  1.95e-82  8.28e-79
6279  FBgn0034434    104.5     171.8     14.72     0.0856          -3.55  8.85e-42  9.40e-39
6494  FBgn0034736    203.9     334.9     29.38     0.0877          -3.51  6.00e-41  5.88e-38
\end{Soutput}
\end{Schunk}

\item Count the number of genes with significant differential expression at FDR of 10\%: \label{DESeq-common2-start}
\begin{Schunk}
\begin{Sinput}
> table( res$padj < 0.1 )
\end{Sinput}
\begin{Soutput}
FALSE  TRUE 
11861   885 
\end{Soutput}
\end{Schunk}








\item Create persistent storage of results using, for example, a \fileformat{CSV} file: \label{persistent_storage_DESeq}


\begin{Schunk}
\begin{Sinput}
> write.csv(res, file="res_DESeq.csv")
\end{Sinput}
\end{Schunk}


\item Perform a sanity check by inspecting a histogram of unadjusted $p$-values (see Figure~\ref{fighistp}) for the differential expression results, as follows:
\label{inspectpvalhist} \label{DESeq-common2-end}
\begin{Schunk}
\begin{Sinput}
> hist(res$pval, breaks=100)
\end{Sinput}
\end{Schunk}

\end{enumerate}


\subsection{D. \tool{DESeq} - complex design}

\begin{enumerate}[i)]

\item Follow Step \ref{DEanalysis} C \ref{DESeq-common-start})-\ref{DESeq-common-end}).

\item Calculate the CR adjusted profile likelihood\cite{Cox1987} dispersion estimates relative to the factors specified, developed by McCarthy et al.\cite{McCarthy2012}, according to:

\begin{Schunk}
\begin{Sinput}
> cds = estimateDispersions(cds, method = "pooled-CR",
    modelFormula = count ~ LibraryLayout + condition)
\end{Sinput}
\end{Schunk}



\item Test for differential expression in the GLM setting by fitting both a full model and reduced model (i.\,e., with the factor of interest taken out):

\begin{Schunk}
\begin{Sinput}
> fit1 = fitNbinomGLMs(cds, count ~ LibraryLayout + condition)
> fit0 = fitNbinomGLMs(cds, count ~ LibraryLayout)
\end{Sinput}
\end{Schunk}

\item Using the two fitted models, compute likelihood ratio statistics and associated P-values, as follows:

\begin{Schunk}
\begin{Sinput}
> pval = nbinomGLMTest(fit1, fit0)
\end{Sinput}
\end{Schunk}


\item Adjust the reported $p$ values for multiple testing:

\begin{Schunk}
\begin{Sinput}
> padj = p.adjust(pval, method="BH")
\end{Sinput}
\end{Schunk}

\item Assemble a result table from full model fit and the raw and adjusted
P-values and print the first few up- and down-regulated genes (FDR less than 10\%):

\begin{Schunk}
\begin{Sinput}
> res = cbind(fit1, pval=pval, padj=padj)
> resSig = res[which(res$padj < 0.1),]
> head( resSig[ order(resSig$conditionKD, decreasing=TRUE), ] )
\end{Sinput}
\begin{Soutput}
            (Intercept) LibraryLayoutSINGLE conditionKD deviance converged     pval     padj
FBgn0013696      -70.96              36.829       37.48 5.79e-10      TRUE 3.52e-03 5.51e-02
FBgn0085822       -5.95               4.382        5.14 2.07e+00      TRUE 4.72e-03 7.07e-02
FBgn0085827       -3.96               4.779        5.08 2.89e+00      TRUE 5.60e-03 8.05e-02
FBgn0264344       -2.59               2.506        4.26 6.13e-01      TRUE 3.86e-04 9.17e-03
FBgn0261673        3.53               0.133        3.37 1.39e+00      TRUE 0.00e+00 0.00e+00
FBgn0033065        2.85              -0.421        3.03 4.07e+00      TRUE 8.66e-15 1.53e-12
\end{Soutput}
\begin{Sinput}
> head( resSig[ order(resSig$conditionKD, decreasing=FALSE), ] )
\end{Sinput}
\begin{Soutput}
            (Intercept) LibraryLayoutSINGLE conditionKD deviance converged    pval  padj
FBgn0031923        1.01              1.2985      -32.26     1.30      TRUE 0.00528 0.077
FBgn0085359        6.37              0.5782       -4.62     3.32      TRUE 0.00000 0.000
FBgn0039155       10.16              0.0348       -4.62     3.39      TRUE 0.00000 0.000
FBgn0024288        6.71             -0.4840       -4.55     1.98      TRUE 0.00000 0.000
FBgn0039827        8.79             -0.2272       -4.06     2.87      TRUE 0.00000 0.000
FBgn0034736        8.54             -0.3123       -3.57     2.09      TRUE 0.00000 0.000
\end{Soutput}
\end{Schunk}

\item Follow Step \ref{DEanalysis} C \ref{DESeq-common2-start})-\ref{DESeq-common2-end}).

\end{enumerate}

\Step{As another spot check, point the \tool{IGV} genome browser (with \fileformat{GTF} and \fileformat{BAM} files loaded) to a handful of the top differentially expressed genes and confirm that the counting and differential expression statistics are appropriately represented.}







\section*{TIMING}

Running this protocol on the SRA-downloaded data will take $\sim$10 hours on a machine with eight cores and 8 GB of RAM; with a machine with more cores, mapping of different samples can be run simultaneously. The time is largely spent on quality checks of reads, read alignment and feature counting; computation time for the differential expression analysis is comparatively smaller.  

\noindent Step \ref{qc}, Sequence quality checks, $\sim$2 h \ \\
Step \ref{meta-start}-\ref{meta-end}, Organizing metadata: $\sim$<1 h \ \\
Steps \ref{mapping-start}-\ref{mapping-end}, Read alignment: $\sim$6 h \ \\
Step \ref{counting}, Feature counting: $\sim$3 h \ \\
Step \ref{DEanalysis}, Differential analysis: variable; computational time is often <20 min \ \\

\section*{TROUBLESHOOTING}

Troubleshooting advice can be found in \textbf{Table 1}. 

\noindent \textbf{Table 1. Troubleshooting} \ \\

\noindent \begin{tabular}{|p{1in}| p{1.2in} | p{1.2in} | p{2.8in} |}

\hline
Step(s) & Problem & Possible reason & Solution \\ \hline
\ref{qc},\ref{DEanalysis}~A~\ref{load_edgeR},\ref{DEanalysis}~C~\ref{load_deseq} & An error occurs when loading a \tool{Bioconductor} package & Version mismatch & Make sure the most recent version of \tool{R} is installed; reinstall packages using \Rfunction{biocLite} \\ \hline
\ref{mapping-start}-\ref{mapping-end} & An error occurs while mapping reads to reference genome & Wrong files made available or version mismatch & Carefully check the command submitted, the documentation for the aligner and the setup steps (e.\,g., building an index); check that there is no clash between \tool{bowtie2} and \tool{bowtie2} \\ \hline
\ref{counting} & An error occurs counting features & \fileformat{GTF} format violation & Use an Ensembl \fileformat{GTF} format or 
coerce your file into a compatible format. In particular, verify that each line of type \emph{exon} contains attributes named \emph{gene\_id} and \emph{transcript\_id}, and that their values are correct. \\ \hline
\ref{DEanalysis} & Errors in fitting statistical models or running statistical tests & Wrong inputs, outdated version of software & Ensure versions of \tool{R} and \tool{Bioconductor} packages are up to date and check the command issued; if command is correct and error persists, post a message to \tool{Bioconductor} mailing list (\url{http://bioconductor.org/help/mailing-list/}) following the posting guide (\url{http://bioconductor.org/help/mailing-list/posting-guide/}. \\ \hline
\end{tabular}

\section*{ANTICIPATED RESULTS}

\noindent \textbf{Sequencing quality checks.} Step \ref{qc} results in an \fileformat{HTML} report for all included \fileformat{fastq} files.  Users should inspect these and look for persistence of low quality scores, overrepresentation of adapter sequence and other potential problems.  From these inspections, users may choose to remove low-quality samples, trim ends of reads (e.\,g., using \tool{FASTX}; \url{http://hannonlab.cshl.edu/fastx_toolkit/}) or modify alignment parameters.  Note that a popular non-\tool{Bioconductor} alternative for sequencing quality checks is \tool{FastQC} (\url{http://www.bioinformatics.babraham.ac.uk/projects/fastqc/}).

\noindent \textbf{Feature counting.}  In Step~\ref{counting}, we used \tool{htseq-count} for feature counting.  The output is a \fileformat{count} file (2-columns: identifier, count) for each sample.  Many alternatives exist inside and outside of \tool{Bioconductor} to arrive at a table of counts given \fileformat{BAM} (or \fileformat{SAM}) files and a set of features (e.\,g., from a \fileformat{GTF} file); see Box~\ref{counting-box} for further considerations.  Each cell in the count table will be an integer that indicates how many reads in the sample overlap with the respective feature. Non-informative rows, such as features that are not of interest or those that have low overall counts can be filtered.  Such filtering (so long as it is independent of the test statistic) is typically beneficial for the statistical power of the subsequent differential expression analysis\cite{Bourgon2010}.

\noindent \textbf{``Normalization''.} As different libraries will be sequenced to different depths, the count data are scaled (in the statistical model) so as to be comparable.  The term \emph{normalization} is often used for that, but it should be noted that the raw read counts are not actually altered\cite{Robinson2010}. By default, \tool{edgeR} uses the number of mapped reads (i.\,e., count table column sums) and estimates an additional normalization factor to account for sample-specific effects (e.\,g., diversity)\cite{Robinson2010}; these two factors are combined and used as an \emph{offset} in the NB model.  Analagously, \tool{DESeq} defines a \emph{virtual} reference sample by taking the median of each gene's values across samples, and then computes \emph{size factors} as the median of ratios of each sample to the reference sample. Generally, the ratios of the size factors should roughly match the ratios
of the library sizes.  Dividing each column of the count table by the corresponding size factor yields normalized count values, which can be scaled to give a \emph{counts per million} interpretation (see also \tool{edgeR}'s \Rfunction{cpm} function).  From an M (log-ratio) versus A (log-expression-strength) plot, count datasets typically show a (left-facing) trombone shape, reflecting the higher variability of log-ratios at lower counts (See Figure~\ref{fig_plotSmear_plotMA}).  In addition, points will typically be centered around a log-ratio of 0 if the normalization factors are calculated appropriately, although this is just a general guide.

\begin{figure}
\centering
\includegraphics[width=\textwidth]{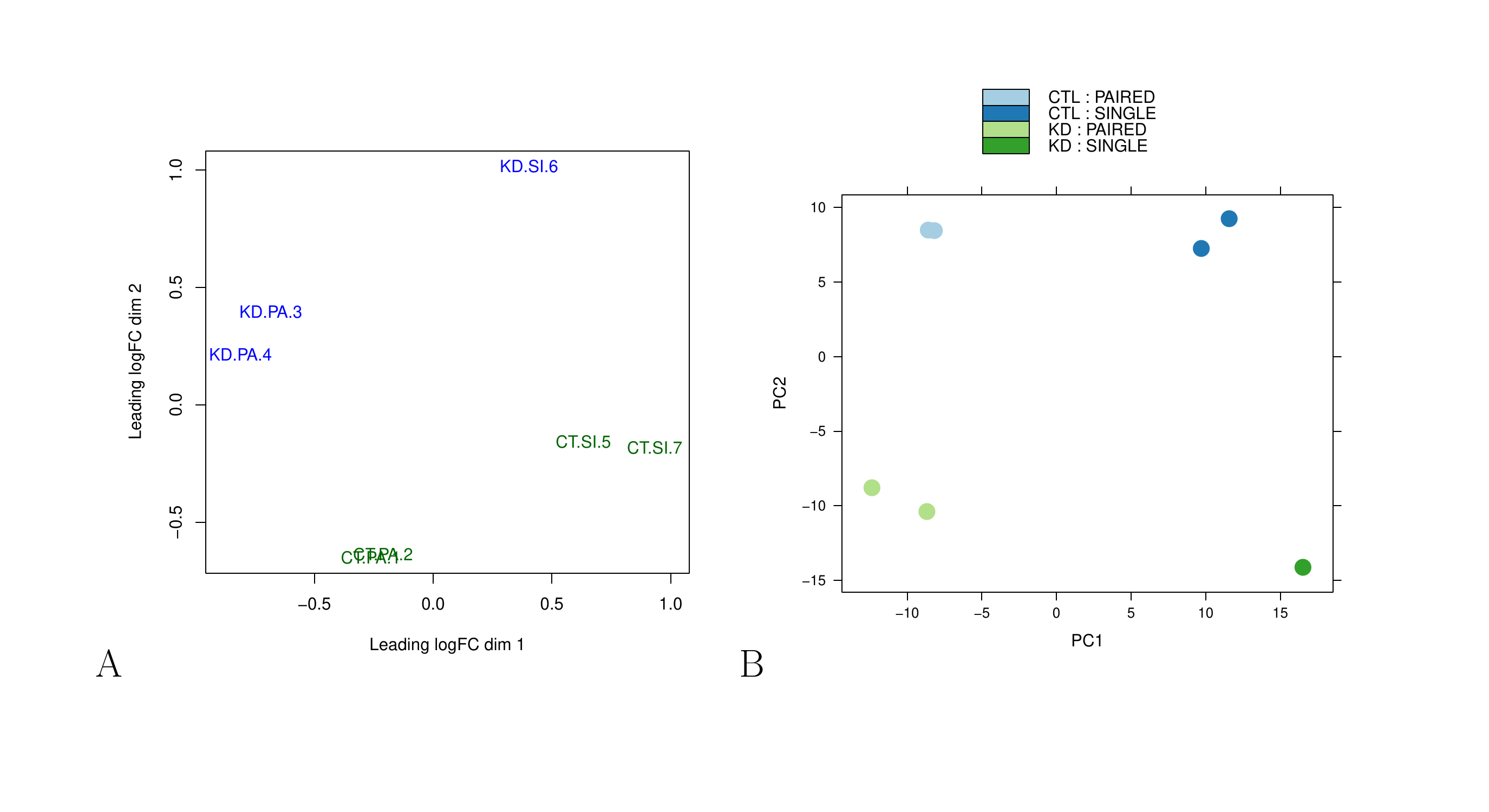}
\caption{Plots of sample relations.  A. Using a count-specific distance measure, \tool{edgeR}'s \Rfunction{plotMDS} produces a multidimensional scaling plot showing the relationship between all pairs of samples.  B. \tool{DESeq}'s \Rfunction{plotPCA} makes a principal component plot of vst-transformed count data.}
\label{fig_MDS_PCA}
\end{figure}

\noindent \textbf{Sample relations.}  The quality of the sequencing reactions (Step~\ref{qc}) themselves are only part of the quality assessment procedure.  In Steps~\ref{DEanalysis}~A~\ref{sampleRelations-edgeR} or \ref{DEanalysis}~C~\ref{sampleRelations-deseq}, a ``fitness for use''\cite{Cappiello2004} check is performed (relative to the biological question of interest) on the count data before statistical modeling.  \tool{edgeR} adopts a straightforward approach that compares the relationship between all pairs of samples, using a count-specific pairwise distance measure (i.\,e., biological coefficient of variation) and an MDS plot for visualization (Figure~\ref{fig_MDS_PCA}A).  Analagously, \tool{DESeq} performs a variance-stabilizing transformation and explores sample relationships using a PCA plot (Figure~\ref{fig_MDS_PCA}B).  In either case, the analysis for the current data set highlights that library type (single-end or paired-end) has a systematic effect on the read counts and provides an example of a data-driven modeling decision: here, a GLM-based analysis that accounts for the (assumed linear) effect of library type jointly with the biological factor of interest (i.\,e., Knockdown versus Control) is recommended.  In general, users should be conscious that the degree of variability between biological replicates (e.\,g., in an MDS or PCA plot) will ultimately impact  the calling of differential expression.  For example, a single outlying sample may drive increased dispersion estimates and compromise the discovery of differentially expressed features.  No general prescription is available for when and whether to delete outlying samples.

\noindent \textbf{Dispersion estimation.} As mentioned above, getting good estimates of the dispersion parameter is critical to the inference of differential expression. For simple designs, \tool{edgeR} uses the quantile-adjusted conditional maximum (weighted) likelihood estimator\cite{Robinson2007,Robinson2008}, whereas \tool{DESeq} uses a method-of-moments estimator\cite{Anders2010}.  For complex designs, the dispersion estimates are made relative to the design matrix, using the CR adjusted likelihood\cite{Cox1987,McCarthy2012}; both \tool{DESeq} and \tool{edgeR} use this estimator.  \tool{edgeR}'s estimates are always moderated toward a common trend, whereas \tool{DESeq} chooses the maximum of the individual estimate and a smooth fit (dispersion versus mean) over all genes.  A wide range of dispersion-mean relationships exist in RNA-seq data, as viewed by \tool{edgeR}'s \Rfunction{plotBCV} or \tool{DESeq}'s \Rfunction{plotDispEsts}; case studies with further details are presented in both \tool{edgeR}'s and \tool{DESeq}'s user guides.

\noindent \textbf{Differential expression analysis.} \tool{DESeq} and \tool{edgeR} differ slightly in the format of results outputted, but each contain columns for (log) fold change, (log) counts-per-million (or mean by condition), likelihood ratio statistic (for GLM-based analyses), as well as raw and adjusted P-values.  By default, P-values are adjusted for multiple testing using the Benjamini-Hochberg\cite{Benjamini1995} procedure.  If users enter tabular information to accompany the set of features (e.g. annotation information), \tool{edgeR} has a facility to carry feature-level information into the results table.  



%
\noindent \textbf{Post differential analysis sanity checks.}  Figure~\ref{fighistp} (Step~\ref{DEanalysis}~C~\ref{inspectpvalhist}) shows the typical features of a $P$-value histogram resulting from a good data set: a sharp peak at the left side, containing genes with strong differential expression, a ``floor'' of values that are approximately uniform in the interval
$[0,1]$, corresponding to genes that are not differentially expressed
(for which the null hypothesis is true), and a peak at the upper end,
at 1, resulting from discreteness of the Negative Binomial test for genes
with overall low counts. The latter component is often less
pronounced, or even absent, when the likelihood ratio test is
used.  In addition, users should spot check genes called as differentially expressed by loading the sorted \fileformat{BAM} files into a genome browser.






\section*{FIGURE LEGENDS}

\begin{infobox}[The Negative Binomial (NB) model]
\label{NBbox}
The NB model has been shown to be a good fit to RNA-seq data\cite{McCarthy2012}, yet flexible enough to account for biological variability. It provides a powerful framework (e.\,g.\ via generalized linear models; GLMs) for analyzing arbitrarily complex experimental designs.
NB models, as applied to genomic count data, make the assumption that an observation, say $Y_{gj}$ (observed number of reads for gene $g$ and sample $j$), has mean $\mu_{gj}$ and variance $\mu_{gj}+ \phi_{g} \mu_{gj}^2$, where the \emph{dispersion} $\phi_g>0$ represents over-dispersion relative to the Poisson distribution\cite{Robinson2007}. The mean parameters $\mu_{gj}$ depend on the sequencing depth for sample $j$ as well as on the amount of RNA from gene $g$ in the sample. Statistical procedures can be formulated to test for changes in expression level between experimental conditions, possibly adjusting for batch effects or other covariates, and to estimate the log-fold-changes in expression.

The dispersion $ \phi_g$ represents the squared coefficient of variation of the true expression levels between biologically independent RNA samples under the same experimental conditions, and hence $\sqrt{\phi}_g$ is called the \emph{biological coefficient of variation}\cite{McCarthy2012}.

Obtaining stable estimates of the genewise dispersions is critical for reliable statistical testing.  Methods of estimating the genewise dispersion estimators have received considerable attention\cite{Robinson2007,Anders2010,VanDeWiel2013,Wu2012}. 
Unless the number of samples is very large, stable estimation of the dispersion requires some sort of sharing of information between genes. One can average the variability across all genes\cite{Robinson2008}, or fit an global trend to the dispersion\cite{Anders2010} or can seek a more general compromise between individual gene and global dispersion estimators\cite{Robinson2007}.
\end{infobox}

\begin{infobox}[Differences between \tool{DESeq} and \tool{edgeR}]
\label{edgeRdeseqdiffbox}
The two packages described in this protocol, \tool{DESeq} and \tool{edgeR}, have similar strategies to perform differential analysis for count data.  However, they differ in a few important areas.  First, their ``look-and-feel'' differs.  For users of the widely-used \tool{limma} package\cite{Smyth2005} (for analysis of microarray data), the data structures and steps in \tool{edgeR} follow analagously.  The packages differ in their default normalization: \tool{edgeR} uses the trimmed mean of M-values\cite{Robinson2010}, while \tool{DESeq} uses a ``relative log expression'' approach by creating a virtual library that every sample is compared against; in practice, the normalization factors are often similar.  Perhaps most critical, the tools differ in the choices made to estimate the dispersion.  \tool{edgeR} moderates feature-level dispersion estimates towards a trended mean according to the dispersion-mean relationship.  In contrast, \tool{DESeq} takes the maximum of the individual dispersion estimates and the dispersion-mean trend.  In practice, this means \tool{DESeq} is less powerful while \tool{edgeR} is more sensitive to outliers.  Recent comparison studies have highlighted that no single method dominates another across all settings\cite{Nookaew2012,Soneson2013,Rapaport2013}.
\end{infobox}

\begin{infobox}[Feature counting]
\label{counting-box}
In principle, counting reads that map to a catalog of features is straightforward.  However, a few subtle decisions need to be made.  For example, how should reads that fall within intronic regions (i.\,e., between two known exons) or beyond the annotated regions be counted?  Ultimately, the answer to this question is guided by the chosen catalog that is presented to the counting software; depending on the protocol used, users should be conscious to include all features that are of interest, such as poly-adenylated RNAs, small RNAs, long intergenic non-coding RNAs and so on.  For simplicity and to avoid problems with mismatching chromosome identifiers and inconsistent coordinate systems, we recommend using the curated \fileformat{FASTA} files and \fileformat{GTF} files from \tool{Ensembl} or the pre-built indices packaged with \fileformat{GTF} files from \url{http://tophat.cbcb.umd.edu/igenomes.html}, when possible.

Statistical inference based on the negative binomial distribution requires raw read counts as input. This is required to correctly model the Poisson component of the sample-to-sample variation. Therefore, it is crucial that \emph{units of evidence} for expression are counted. No prior normalization or other transformation should be applied, including quantities such as RPKM (reads per kilobase model), FPKM (fragments per kilobase model) or otherwise depth-adjusted read counts. Both \tool{DESeq} and \tool{edgeR} internally keep the raw counts and normalization factors separate, as this full information is needed to correctly model the data.  Notably, recent methods to normalize RNA-seq data for sample-specific G+C content effects employ offsets that are presented to the GLM, while maintaining counts on their original scale\cite{Hansen2012,Risso2011}.  

Paired-end reads each represent a single fragment of sequenced DNA, yet (at least) two entries for the fragment will appear in the corresponding \fileformat{BAM} files.  Some simplistic early methods that operated on \fileformat{BAM} files considered these as separate entries, which led to overcounting and would ultimately overstate the significance of differential expression.   

Typically, there will be reads that cannot be uniquely assigned to a gene, either because the read was aligned to multiple locations (multi-reads) or the read's position is annotated as part of several overlapping features. For the purpose of calling differential expression, such reads should be discarded. Otherwise, genuine differential expression of one gene might cause another gene to appear differentially expressed, erroneously, if reads from the first gene are counted for the second due to assignment ambiguity.  In this Protocol, we employ the tool \tool{htseq-count} of the Python package HTSeq using the default \emph{union} counting mode; more details can be found at \url{http://www-huber.embl.de/users/anders/HTSeq/doc/count.html}.  In addition, \tool{Bioconductor} now offers various facilities for feature counting, including \Rfunction{easyRNASeq} in the \tool{easyRNASeq} package\cite{Delhomme2012},  \Rfunction{summarizeOverlaps} function in the \tool{GenomicRanges} package and \Rfunction{qCount} in the \tool{QuasR} (\url{http://www.bioconductor.org/packages/release/bioc/html/QuasR.html}) package.
\end{infobox}

\begin{infobox}[Software versions]
\label{versionbox}
The original of this document was produced with Sweave\cite{Leisch2002} 
using the following versions of \tool{R} and its packages:

\begin{Schunk}
\begin{Sinput}
> sessionInfo()
\end{Sinput}
\begin{Soutput}
R version 3.0.0 (2013-04-03)
Platform: x86_64-unknown-linux-gnu (64-bit)

locale:
 [1] LC_CTYPE=en_CA.UTF-8       LC_NUMERIC=C               LC_TIME=en_CA.UTF-8       
 [4] LC_COLLATE=en_CA.UTF-8     LC_MONETARY=en_CA.UTF-8    LC_MESSAGES=en_CA.UTF-8   
 [7] LC_PAPER=C                 LC_NAME=C                  LC_ADDRESS=C              
[10] LC_TELEPHONE=C             LC_MEASUREMENT=en_CA.UTF-8 LC_IDENTIFICATION=C       

attached base packages:
[1] parallel  stats     graphics  grDevices utils     datasets  methods   base     

other attached packages:
 [1] DESeq_1.12.0         locfit_1.5-9.1       Biobase_2.20.0       edgeR_3.2.3         
 [5] limma_3.16.2         ShortRead_1.18.0     latticeExtra_0.6-24  RColorBrewer_1.0-5  
 [9] Rsamtools_1.12.3     lattice_0.20-15      Biostrings_2.28.0    GenomicRanges_1.12.4
[13] IRanges_1.18.1       BiocGenerics_0.6.0   cacheSweave_0.6-1    stashR_0.3-5        
[17] filehash_2.2-1      

loaded via a namespace (and not attached):
 [1] annotate_1.38.0      AnnotationDbi_1.22.5 bitops_1.0-5         DBI_0.2-7           
 [5] digest_0.6.3         genefilter_1.42.0    geneplotter_1.38.0   grid_3.0.0          
 [9] hwriter_1.3          RSQLite_0.11.3       splines_3.0.0        stats4_3.0.0        
[13] survival_2.37-4      tools_3.0.0          XML_3.96-1.1         xtable_1.7-1        
[17] zlibbioc_1.6.0      
\end{Soutput}
\end{Schunk}

The versions of software packages used can be captured with the following commands:

\begin{Schunk}
\begin{Sinput}
> system("bowtie2 --version | grep align", intern=TRUE)
\end{Sinput}
\begin{Soutput}
[1] "/usr/local/software/bowtie2-2.1.0/bowtie2-align version 2.1.0"
\end{Soutput}
\end{Schunk}

\begin{Schunk}
\begin{Sinput}
> system("tophat --version", intern=TRUE)
\end{Sinput}
\begin{Soutput}
[1] "TopHat v2.0.8"
\end{Soutput}
\end{Schunk}

\begin{Schunk}
\begin{Sinput}
> system("htseq-count | grep version", intern=TRUE)
\end{Sinput}
\begin{Soutput}
[1] "General Public License v3. Part of the 'HTSeq' framework, version 0.5.3p9."
\end{Soutput}
\end{Schunk}

\begin{Schunk}
\begin{Sinput}
> system("samtools 2>&1 | grep Version", intern=TRUE)
\end{Sinput}
\begin{Soutput}
[1] "Version: 0.1.18 (r982:295)"
\end{Soutput}
\end{Schunk}
\end{infobox}

\section*{SUPPLEMENTARY MATERIAL}

\noindent \textbf{Supplementary File 1.}  This file is a compressed archive with the following files:  the intermediate \fileformat{COUNT} files used, a count table used in the statistical analysis, the metadata table and the original ``SraRunInfo'' \fileformat{CSV} file that was downloaded from the NCBI's Short Read Archive.

\begin{addendum}
\item The authors wish to thank Xiaobei Zhou for comparing counting methods, Olga Nikolayeva for feedback on an earlier version of the manuscript and members of the ECCB Workshop (Basel, September 2012) for their feedback. DJM is funded by the General Sir John Monash Foundation, Australia.  MDR wishes to acknowledge funding from the University of Zurich's Research Priority Program in Systems Biology and Functional Genomics and SNSF Project Grant (143883).  SA, WH and MDR acknowledge funding from the European Commission through the 7$^{th}$ Framework Collaborative Project RADIANT (Grant Agreement Number: 305626).  
\item[Competing Interests] The authors declare that they have no
competing financial interests.
\item[Author Contributions] SA and WH are authors of the \tool{DESeq} package.  DJM, YC, GKS and MDR are authors of the \tool{edgeR} package.  SA, MO, WH and MDR initiated the protocol format, based on the ECCB 2012 Workshop.  SA and MDR wrote the first draft and additions were made from all authors.
\end{addendum}


\section*{REFERENCES}

\bibliography{protocol}
\bibliographystyle{naturemag}

\end{document}